# Designing thermoplasmonic polymersomes for photothermal therapy

*Valentino Barbieri[1,2,3], Javier González-Colsa[4], Diana Matias[1,2,5], Aroa Duro-Castano[1], Anshu Thapa[1], Lorena Ruiz-Perez[1,2,3], Pablo Albella[4], Giorgio Volpe[1,]\* and Giuseppe Battaglia[1,2,3,6]\**

[1]Department of Chemistry, 20 Gordon Street, University College London, WC1H 0AJ London, United Kingdom.

[2]Molecular Bionics Group, Institute for Bioengineering of Catalunya (IBEC), The Barcelona Institute of Science and Technology (BIST) Barcelona, Spain.

[3]Biomedical Research Networking Center in Bioengineering, Biomaterials, and Nanomedicine (CIBER-BBN), Barcelona, Spain.

[4]Group of Optics, Department of Applied Physics, University of Cantabria, 39005, Santander, Spain.

[5]Instituto de Medicina Molecular João Lobo Antunes (iMM), Lisbon, Portugal.

[6]Catalan Institution for Research and Advanced Studies (ICREA), Barcelona, Spain.

Polymersomes, vesicles self-assembled from amphiphilic polymers, are promising nanocarriers for the targeted intracellular delivery of therapeutics. Integrating inorganic light-absorbing materials with plasmonic properties, such as gold, into their membrane by in situ synthesis is a stepping stone to enable their use for photothermal therapy. Yet, it still needs to be determined whether the in situ synthesis of gold can produce polymersomes with thermoplasmonic properties without altering their morphology, stability, and nanocarrier functionality. Here we demonstrate that small gold nanoparticles can be controllably nucleated within biocompatible block-copolymer membranes to design hybrid polymersomes with a noteworthy thermoplasmonic response. The cumulative absorption of individual 2 nm gold nanoparticles can induce temperature increases in 10 K in dilute suspensions of hybrid polymersomes upon laser illumination.
Furthermore, we develop a theoretical model to rationalize our observations and predict the thermoplasmonic response of our hybrid polymersomes. We finally demonstrate in vitro photothermal therapy of cancer cells, enhanced by the receptor-mediated endocytosis of our hybrid polymersomes. We envision that our nanotechnological platform can be translated to phenotypic cell targeting to precisely deliver effective photothermal agents.

1. Introduction

Gold nanoparticles (AuNPs) have attracted significant attention in nanomedicine due to their ability to efficiently respond to light stimulation and induce light-to-heat conversion at the micro/nanoscale.[1] These distinct characteristics come from the excitation of localized surface plasmons resonances (LSPRs), collective oscillations of conduction electrons on the nanoparticle's surface. [2] At the LSPR frequency, the collective motion of the free electrons is enhanced, leading to an increment of resistive losses that ultimately causes thermal energy generation.[3] Harnessing plasmonic resonances enables applications in various areas of nanomedicine, including biosensing, imaging, and photothermal therapy (PTT). [1,4,5] The latter is a promising non-invasive approach for treating solid tumors relying on laser-induced heat generation to destroy cancer cells selectively.[6] Since hyperthermia conditions require colocalizing both the light stimulus and the AuNPs, PTT offers high precision and minimizes the typical adverse effects commonly associated with traditional cancer treatments.[5] However, AuNPs are colloidally unstable in biological environments, where they often aggregate and interact with the host immune system.[4,7,8] The surface of AuNPs can be functionalized to improve their colloidal stability and extend their bioavailability. Such seemingly inert surface modifications can cause undesired accumulation in some organs and induce toxicity.[9–11] Surface functionalization with polymers can also induce the assembly of AuNPs into more complex structures, including vesicles.[12–14] Yet, these approaches require the lengthy optimization of complex architectures and assembly procedures that can complicate their scalable manufacturing and clinical translation.[15–17]

A workaround is to encapsulate AuNPs into more established nanocarriers such as liposomes or polymer nanoparticles. Among these structures, polymersomes – synthetic vesicles made of amphiphilic block copolymers – comprise optimal attributes for targeted delivery. Such attributes include enhanced colloidal stability, bioavailability, and high chemical versatility. [18,19] AuNPs have thus been encapsulated within polymersome membranes to create hybrid vesicles. Such systems have mainly been produced by the encapsulation of preformed AuNPs during the self-assembly of the polymersome.[20–23] Alternatively, a few studies have reported the in situ synthesis of AuNPs within the polymersome membrane as imaging contrast agents[24] or to control drug release profiles.[25] However, it is unclear whether hybrid polymersomes fabricated via the in situ synthesis of gold can support tangible thermoplasmonic properties due to the small size of the AuNPs supported by the membrane.[26,27]

In this work, we present functional thermoplasmonic polymersomes produced via the in situ synthesis of 2 nm AuNPs within the membranes of pH-sensitive poly[(2-methacryloyl)ethyl phosphorylcholine]$_{25}$-*block*-poly[2-(diisopropylamino)ethyl methacrylate]$_{80}$ (PMPC–PDPA) polymersomes. Previous studies showed that PMPC-PDPA polymersomes without gold can effectively transport various types of cargo into cells through receptor-specific targeting. These cargoes include nucleic acids, small molecules, proteins, and even small nanoparticles.[28–30] Now, after characterizing the morphology of the newly synthesized hybrid polymersomes, we demonstrate their thermoplasmonic properties. In addition, we develop a theoretical model to explain the structure-function relationship in the photothermal response of our system. Finally, we show that their internalization into cancer cells can maximize their plasmonic response for in vitro photothermal therapy.

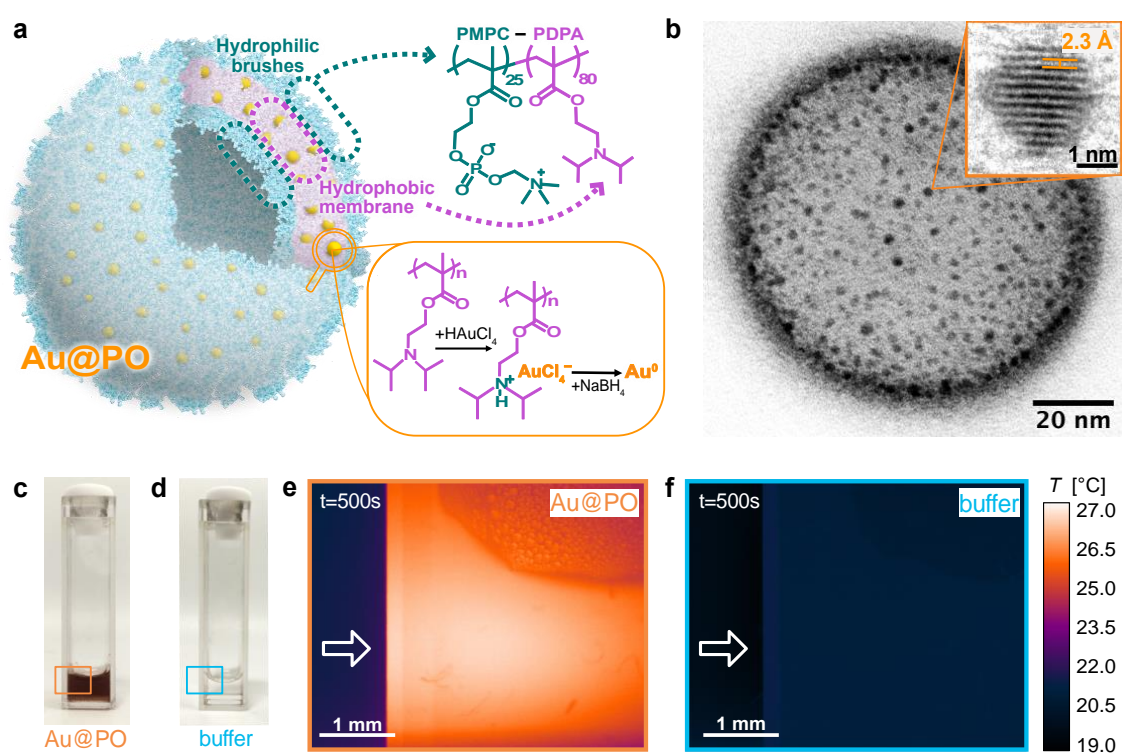

**Figure 1. The design concept of thermoplasmonic hybrid polymersomes.** a) Schematic of a hybrid polymersome (Au@PO). Amphiphilic PMPC–PDPA chains self-assemble into polymersomes in aqueous environments. Gold nanoparticles (AuNPs) are embedded in the PDPA membrane following an in situ nucleation reaction (inset). The acidic gold precursor

(HAuCl$_4$) protonates the PDPA membrane, which acts as a capping agent for reducing gold chloride ions (AuCl$_4^-$) to metallic Au$^0$. b) Representative transmission electron microscopy (TEM) image of a hybrid polymersome. The embedded AuNPs appear as dark dots. The TEM specimen was positively stained with phosphotungstic acid. Inset: high-resolution TEM image of a single gold nanoparticle showing the spacing between the Au(111) lattice planes. c-d) Photographs of two cuvettes containing (c) a dense aqueous suspension of hybrid polymersomes in phosphate buffer solution (6×10$^{12}$ cm$^{-3}$) and (d) the buffer solution only. e-f) Infrared images of the areas highlighted by the respective squares in c-d showing the temperature increase from (e) the Au@PO sample in c and (f) the buffer control in d upon irradiation with a 532 nm laser, power density 0.128 mW µm$^{-2}$. The arrows indicate the point of incidence of the laser beam.

## 2. Results and discussion

### 2.1. Design of thermoplasmonic polymersomes

We fabricated biocompatible hybrid polymersomes made of PMPC–PDPA copolymers embedding randomly distributed AuNPs within the hydrophobic PDPA membrane (**Figure 1a**). This amphiphilic block-copolymer was selected because of its two blocks' physicochemical properties, making it ideal for both the reaction and the vesicles' intended nanomedical application. First, the zwitterionic PMPC groups forming the hydrophilic brushes are known for their antifouling properties.[31–33] Hence, PMPC can minimize the formation of a 'protein corona' around the polymersomes when immersed in crowded biological environments.[34] Moreover, we have demonstrated that the PMPC motif targets three cell receptors, SRB1, CD36, and CD8; thus, PMPC enables targeted delivery to cancer cells or myeloid cells.[35–37] Lastly, the PDPA hydrophobic block in the membrane contains pH-sensitive tertiary amine groups capable of coordinating gold ions and templating their crystallization in situ.[25] In our case, we produced PMPC–PDPA polymersomes by the 'solvent switch' protocol (Methods), which involves the bottom-up assembly of amphiphilic polymer chains when a good solvent is replaced with an aqueous buffer. Depending on the composition of the buffer, polymersome formulations with average diameters of 45, 60, or 100 nm could be obtained. We then proceeded to the in situ synthesis of the AuNPs (Methods). The first step involves the partial protonation of PDPA membranes by the acidic gold precursor HAuCl$_4$. We controlled the pH using a phosphate saline buffer, PBS 0.1 M, and pH 7.4 to avoid complete protonation and disassembly of the polymersomes. The positively charged polymersome membrane attracts the AuCl$_4^-$ ions, which diffuse to the membrane and interact with the amine in the PDPA block. After adding the NaBH$_4$ reducing agent, the nucleation of metallic Au-seeds occurs; hence, the AuNPs growth is templated onto the polymer chains. [24]

Transmission electron microscopy (TEM) imaging confirmed the successful outcome of the reaction (**Figure 1b**). To visualize the polymer assembly, we stained the specimen with phosphotungstic acid (PTA), which acts as positive staining on PMPC–PDPA (Methods).[38] The resulting image indicates a 'chocolate-chip cookie' morphology, in which the AuNPs appear as dark dots included within the organic matrix. Furthermore, we performed high-resolution TEM imaging (Figure 1b, inset) on the embedded AuNPs, to resolve the lattice planes and assign a crystallographic structure. The measured interplanar spacing of (0.23±0.01) nm is in good agreement with the reported value for Au(111) planes in the face-centered cubic (fcc) unit cell adopted by plasmonic sub-5 nm AuNPs. [39,40]

A simple visual inspection of hybrid polymersomes dispersed in buffer suggests their ability to absorb visible light, resulting in a deep brown color (**Figure 1c**). Conversely, the buffer alone does not present any coloration, as expected (**Figure 1d**). Infrared imaging (Methods) confirms that the hybrid polymersome dispersion shows the ability to dissipate the absorbed radiation in the form of heat when illuminated with a 532 nm laser (**Video S1**). **Figure 1e-f** show the final temperature increment reached by the dispersion and the control, respectively, after 500 s of illumination. While the hybrid polymersome dispersion (Figure 1e) generated a net increase in temperature at the laser incidence spot, the buffer alone did not present any detectable deviation from room temperature.

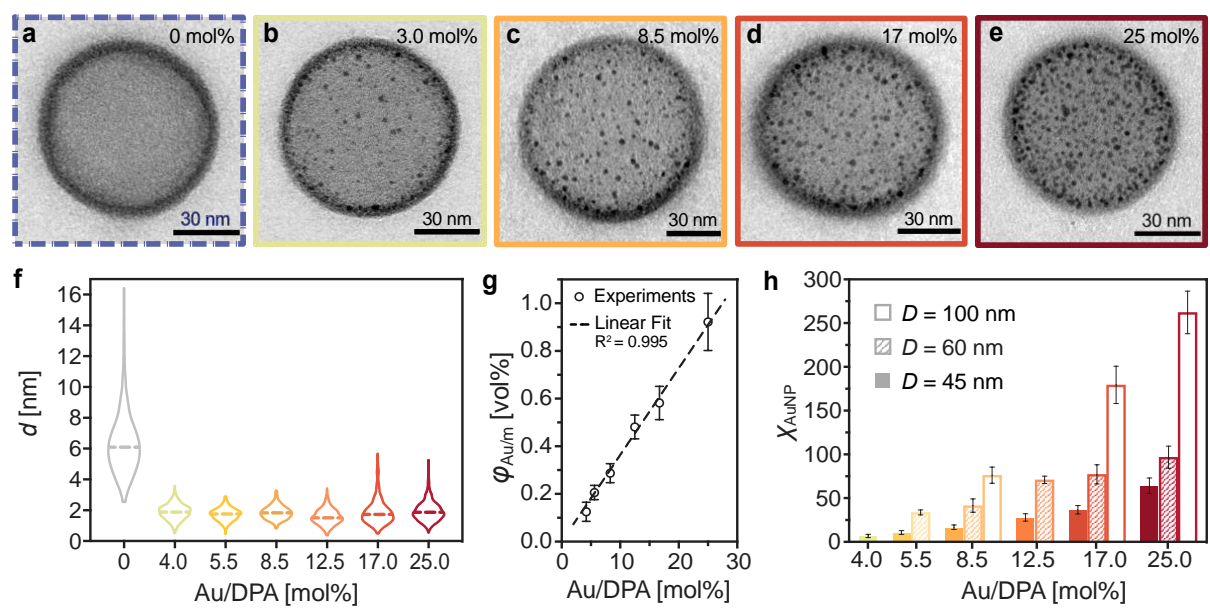

**Figure 2. Morphological characterization of hybrid polymersomes.** a-e) Representative transmission electron micrographs of (a) pristine polymersome and (b-d) hybrid polymersomes synthesized at increasing gold precursor-to-monomer ratios (Au/DPA): (b) 3 mol%, (c) 8.5 mol%, (d) 17% and (e) 25 mol%. f) Violin plots showing that the diameter *d* of the gold nanoparticles (AuNPs) embedded in the polymersome membrane is relatively constant around 2 nm and thus unaffected by the initial Au/DPA ratio. The 0 mol% ratio corresponds to free AuNPs synthesized in the absence of polymersomes and is shown for reference. Each violin plot represents the distributions of at least 450 AuNPs as measured from TEM images, the dashed lines mark the medians. g) Linear relationship between the measured final volume fraction of gold within the polymersome membrane $\varphi_{Au/m}$ and the initial Au/DPA ratio fed into the reaction. The error bars around data points represent the standard deviation from three repeats per sample. h) The gold nanoparticle loading density ($\chi_{AuNP}$) is directly proportional to the initial Au/DPA ratio and the mean polymersome diameter *D*. Columns report mean values; error bars represent one standard deviation between at least three repeated measures per sample.

2.2. Morphological characterization

To explore the gold-loading capabilities of our system, we synthesized hybrid polymersomes at different HAuCl$_4$ concentrations, expressed as molar Au/DPA percentages (mol%), where DPA indicates the repeating unit of the PDPA block (Methods). Within the 3-25 mol% concentration range, the reaction produced hybrid polymersomes with substantially unchanged size distributions and colloidal stabilities compared to the pristine polymersome dispersions, as confirmed by dynamic light scattering (DLS) measurements (**Figure S1**). Concentrations higher than 25 mol% did not yield stable systems in the pH buffers used for the synthesis. Introducing any more HAuCl$_4$ caused the pH to drop too far below the pKa of the polymer, leading to severe alteration or even complete polymersome disassembly.[41] The TEM images of hybrid polymersomes shown in **Figure 2a-e** confirm that the polymersomes retained the spherical morphology upon increased Au concentration, further confirming the DLS sizing (Figure S1). TEM examination shows that the final size of AuNPs within the membranes remains constant, whereas their density increases with the concentration of HAuCl$_4$ used in the synthesis reaction. We further confirmed this observation by manually sizing over 450 particles from images of different hybrid polymersomes in each formulation (violin plots in **Figure 2f**). The diameters of the embedded AuNPs remained narrowly distributed around the mean value of (1.9 ± 0.5) nm as the Au/DPA molar ratio was increased, supporting the fact that the AuNP growth is templated within the polymersome membrane. Conversely, when AuNPs are synthesized under the same conditions but in the absence of the polymersomes (Au/DPA = 0 mol% in Figure 2f), significantly larger and more broadly distributed diameters could be measured after the reduction with $NaBH_4$. The bare nanoparticles showed colloidal instability and precipitated soon after the synthesis.

We proceeded to investigate the nucleation pattern of AuNPs within membranes as the Au/DPA ratio grows. At low concentrations, TEM revealed that some polymersomes remained empty as others started to be decorated (**Figure S2a-b**). Above 12.5 mol% of Au/DPA, all polymersomes were found to be hybrid instead (**Figure S2c-d**). To characterize the Au loading more quantitatively, we measured the concentration of Au and PMPC–PDPA in our formulations by microwave-assisted plasma atomic emission spectroscopy (MP-AES) and high-performance liquid chromatography (HPLC), respectively (Methods). We then combined the concentration data with morphological information to build an Au-loading model, as detailed in the SI. Such model was used to calculate the volume ratio of Au in the membrane, $\varphi_{Au/m}$, defined as the ratio between the volume concentrations of Au and the membrane-forming PDPA phases. **Figure 2g** highlights the linear relationship between the volume ratio $\varphi_{Au/m}$ and the input Au/DPA molar percentage fed into the reaction. The plot contains data from 24 different formulations that were synthesized from five different polymersome batches characterized by three average sizes. The observed linearity over such a wide variety of formulations indicates the high control over the final hybrid polymersome morphology afforded by the in situ reaction. The total volume of gold embedded in the membrane of our hybrid polymersomes can indeed be determined in the reaction design stage just by acting on a single variable, i.e., the $HAuCl_4$ precursor concentration.

As a visually descriptive loading parameter, we defined the AuNPs loading density $\chi_{AuNP}$, which indicates the average number of AuNPs per polymersome. Because of the size uniformity of the AuNPs, $\chi_{AuNP}$ displays a similar direct proportionality on the precursor concentration as that observed for the volume ratio $\varphi_{Au/m}$ (**Figure 2h**). Since the number of nucleation sites in the membrane increases with the size of the polymersomes, $\chi_{AuNP}$ also grows together with the mean polymersome diameter *D*. Once again, we emphasize the versatility of our method. The independent control of the polymersomes and AuNPs' morphologies in separate stages of the fabrication allows to formulate a wide portfolio of stable hybrid polymersomes with cargos ranging from a few up to hundreds of AuNPs per polymersome. We finally assessed the long-term stability of the hybrid polymersome formulations by repeating the morphological characterization after one year of storage at 4°C. Both the pristine and hybrid polymersomes exhibited outstanding stability over time, as confirmed by the DLS analysis in **Figure S3a**. The size distributions remained unchanged after one year for all three tested formulations, with the curves shifting by less than the overall uncertainty of the measure (95% confidence interval of the mean). The TEM images in **Figure S3b** show that the spherical shape of the polymersomes and both the size and location of embedded AuNP were also unaffected by the long-term storage.

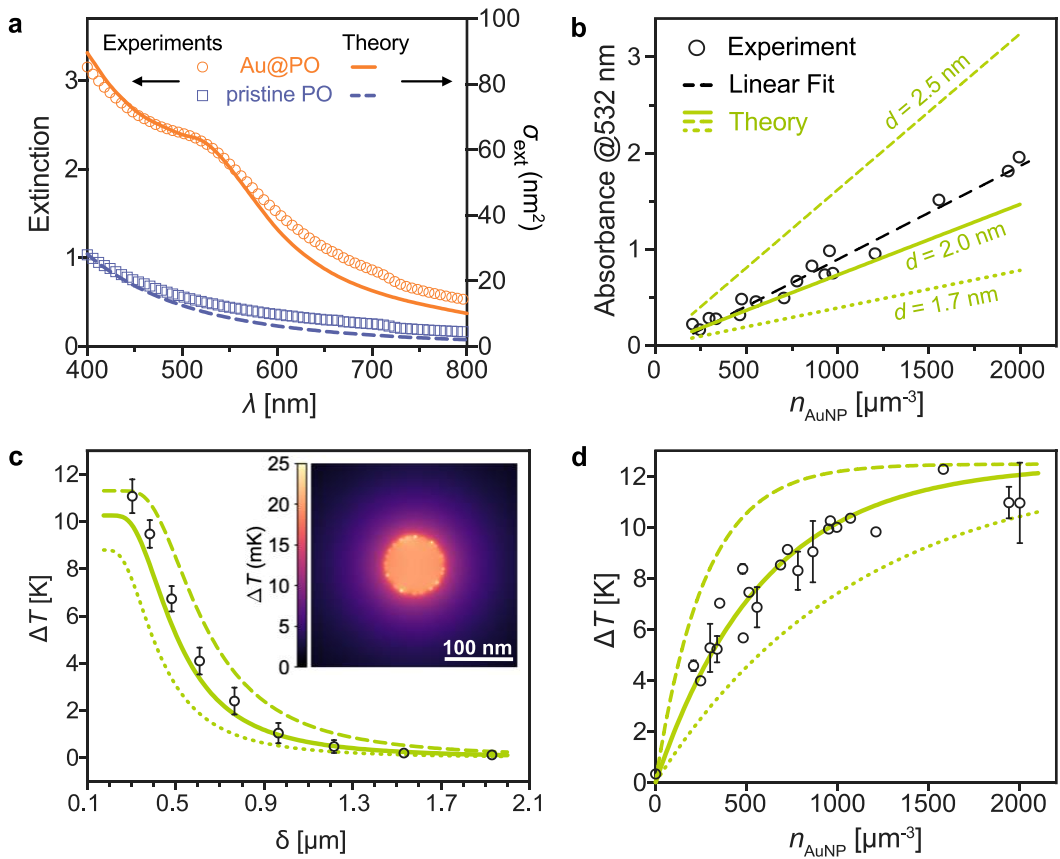

**Figure 3. Thermoplasmonic properties of hybrid polymersomes.** a) Experimental extinction spectra of a hybrid polymersome suspension (Au@POs, orange circles) showing localized plasmon resonance centered around $\lambda \approx 518$ nm. The spectrum of pristine polymersomes (blue squares) is shown for reference. The theoretical extinction cross-sections $\sigma_{ext}$ (lines) qualitatively reproduce the respective experimental spectra. b) The optical absorbance of hybrid polymersomes at 532 nm depends linearly on the number density of AuNPs, $n_{AuNP}$ (circles). A linear fit (black dashed line) allows estimation of the experimental absorption cross-section of an individual AuNP ($\sigma_{abs} = 0.22$ nm$^2$). Predictions calculated with theoretical absorption cross-sections are shown as green lines for the case of 1.7 nm (dotted), 2 nm (solid) and 2.5 nm (dashed) AuNPs, respectively taken as the center, upper and lower

limit of the confidence interval. c) Experimental steady-state temperature increments $\Delta T$ (circles) generated by the exposure of a progressively diluted hybrid polymersome suspension to a 532 nm laser (incident laser intensity $I = 0.128$ mW µm$^{-2}$) as a function of the mean inter-polymersome spacing, $\delta$. Data points are the mean of three measurements, error bars represent one standard deviation. Inset: Example of simulated temperature map for a single 100 nm hybrid polymersome irradiated with the same incident laser intensity $I$. d) Experimental $\Delta T$ (circles) generated by the laser exposure of hybrid polymersome dispersions as a function of the AuNPs number density $n_{\text{AuNP}}$. Data points are the mean of two measurements, error bars represent one standard deviation. Our theoretical model accurately describes the photothermal response of our suspensions, as demonstrated by the green lines in c and d, plotted for the central value (solid), upper limit (dashed) and lower limit (dotted) of the absorption cross-section confidence interval as in b.

2.3. Thermoplasmonic characterization

As demonstrated in the proof-of-concept experiment of **Figure 1e-f**, despite the small sizes of the AuNPs embedded in their membrane, our hybrid polymersomes feature a significant photothermal response. The origin of this effect is the enhanced absorption of photons at the LSPR wavelength. The extinction spectrum of a representative hybrid polymersome suspension is reported in **Figure 3a** (orange circles). While the extinction of pristine polymersomes (blue squares) monotonically decreases as the incident wavelength increases across the visible and near-infrared spectral regions, the LSPR appears in the spectrum of hybrid polymersomes as a broad shoulder feature centered around 518 nm. The flattened and broad shape of the extinction curve around the LSPR is typical of ultrasmall (< 5 nm) AuNPs and is caused by the damping of the electron at these high levels of confinement. [2,42] The shape of the experimental spectra can be accurately reproduced by a simple linear combination of the light-matter interaction phenomena in the two phases of the hybrid system, as calculated through Mie theory (Methods). In particular, the theory suggests that scattering is negligible for AuNPs and absorption is negligible in polymersomes, giving

$$\sigma_{\text{ext}} = \chi_{\text{AuNP}}\, \sigma_{\text{abs}} + \sigma_{\text{sca}} \qquad (1)$$

where $\sigma_{\text{ext}}$ represents the extinction cross-sections of the hybrid polymersome, $\sigma_{\text{abs}}$ the absorption cross-section of a single 2 nm nanoparticle, and $\sigma_{\text{sca}}$ the scattering cross-section of an isolated polymersome having the average diameter of the sample population. The resulting theoretical extinction cross-sections are represented as lines on the right axis of the plot in Figure 3a. Our observation of an LSPR feature in hybrid polymersomes produced by the in situ synthesis differs from the findings in the work of Zhu et al.[27], who suggested that the absence of such a feature could result from the successful immobilization of the AuNPs in the membrane. Our findings demonstrate, instead, that plasmonic properties are compatible with the in situ growth of small nanoparticles embedded in the vesicle membranes. In addition, **Figure S4** shows that the prominence of the LSPR increases with the volume fraction of Au in the membrane. As shown in **Equation 1**, polymersomes account entirely and exclusively for the scattering contribution; therefore, the experimental absorption spectra of hybrid polymersomes can be obtained, in first approximation, by subtracting the extinction spectra of the relative pristine polymersomes from each spectrum. By doing so, we isolated and quantified the cumulative absorbance of AuNPs in all the hybrid polymersome formulations, which causes heat generation from our structures. In **Figure 3b** we show that the absorbance near the LSPR wavelength – here at 532 nm, as used in the thermoplasmonic heating measurements – grows linearly with the concentration of AuNPs in the sample, expressed as number density per unit volume ($n_{\text{AuNP}}$). Such linearity guarantees that the concentrations in our samples are within the validity range of the Beer-Lambert law. Therefore, the relation $A = \left(\frac{\sigma_{\text{abs}} l}{\ln 10}\right) n_{\text{AuNP}}$, where $A$ stands for the absorbance and $l$ for the optical cell thickness, can be applied to extract the experimental absorption cross-section from the gradient of the best-fit line. This yields a value of $(0.22 \pm 0.02)$ nm$^2$, which is slightly higher than the theoretical value calculated for 2 nm particles (solid green line) via extrapolation of the semi-empirical model proposed by Karimi et al.[42] Yet, this value falls within the limiting theoretical cross-sections that are acceptable within our particle size distribution and the validity of Karimi's model.[42]

After describing the optical absorption in hybrid polymersomes and its dependence on the concentration of AuNPs, we investigate the resulting heat generation. We used an internally built and calibrated resistance thermal probe to precisely collect the temperature evolution at the center of a quartz cuvette containing the sample under illumination with a 532 nm laser (Methods), as displayed in the scheme in **Figure S5a**. Some typical temperature-time traces are reported in **Figure S5b**. The recorded temperature evolution results from the heat generation by the hybrid polymersomes in the illuminated volume, followed by heat transfer to the surrounding environment through convection and conduction phenomena. As a result, the temperature in the vicinities of the focal spot grows sharply in the first transient, plateauing as the heat is transferred to the surroundings and the whole sample approaches thermal equilibrium after 700 s of exposure. As expected for thermoplasmonic systems, applying a ramp of increasing laser power, the final equilibrium temperature grows proportionally with the incident power density (Figure S5).

Considering a single formulation and fixing the incident power density to 0.128 mW µm$^{-2}$, we show that the temperature increment, $\Delta T$, decreases from its maximal value of 11 K as the hybrid polymersomes in the suspension are progressively distanced through consecutive dilutions (**Figure 3c**). This occurs because the global temperature measured in the bulk of the suspension is the result of the collective heating of all the hybrid polymersomes in the illuminated volume. The larger the mean nearest-neighbor distance $\delta$ between the heat emitters, the fewer of them will be found in the illuminated volume, and the less the temperature profile developing around each of them will sum to the ones generated by its neighbors. To further investigate this phenomenon, we simulated the thermoplasmonic temperature evolution around a single hybrid polymersome by finite element modeling (Methods). As an example, the inset of Figure 3c shows the temperature field produced by an isolated polymersome at the incident power density of 0.128 mW µm$^{-2}$. The temperature reaches a maximum of 25 mK on the surface of the embedded AuNPs (i.e., the absorbers), from where heat diffuses into the aqueous medium inside and outside the polymersomes. In the polymersome core, due to spatial confinement of the warm liquid, a constant temperature of about 20 mK is maintained, whereas the temperature drops with inverse proportionality to the radial distance in the external surroundings of the polymersomes, reaching values of only 1.7 mK at 200 nm from the center. It is evident, therefore, that the spatial proximity between polymersomes is important for their temperature fields to overlap and build up to a collective effect. Upon continuous exposure, we expect the temperature to equilibrate at longer time scales until a constant value is reached across the whole sample. To calculate this steady state global temperature and theoretically support our observations, we developed a fully analytical theoretical model, the full derivation of which can be found in the SI. AuNPs and polymersomes are modeled as independent absorbers and scatterers (**Equation 1**). Absorption phenomena are considered responsible both for heat generation and, alongside scattering events, for the attenuation of the laser beam as it crosses the sample. After generation, heat is exchanged with the external environment by convective flows along the boundaries and conduction through the cuvette walls. Both phenomena can be accounted for with an overall heat transfer coefficient $\beta$ (see SI). The resulting theoretical $\Delta T$ can be calculated as:

$$\Delta T(\delta) = \frac{P}{\beta} \frac{\chi_{\text{AuNP}} \, \sigma_{\text{abs}}}{\sigma_{\text{ext}}} \left[ 1 - e^{-\sigma_{\text{ext}} l \left(\frac{\kappa}{\delta}\right)^3} \right] \qquad (2)$$

where *P* represents the incident laser power, and *κ* is a constant geometrical factor used to estimate the average inter-polymersome spacing from a random Poisson distribution.[43] One important feature in the equation is that no terms representing the illuminated volume or the spot size are present. This means that the observed linearity of the temperature with the laser intensity observed in Figure S5 only comes from the total power delivered to the sample, regardless of how tightly the laser is focused. We plotted the curve resulting from **Equation 2** in the case of a 100 nm polymersome containing 260 AuNPs ($d = 2$ nm) as a solid green line in Figure 3c, together with the upper (dashed line) and lower (dotted line) limits of the confidence interval defined in Figure 3b. This simple model can accurately and quantitatively describe our data, with all the experimental data points falling within the calculated confidence interval. It is also worth observing that the temperature increment saturates for very concentrated suspensions, for which the inter-polymersome distance is further reduced below 300 nm. This indicates that the maximum photothermal conversion efficiency of the system has been reached.

Our model can be further extended to analyze the evolution of the laser-induced temperature increment in different hybrid polymersome formulations, regardless of their average size, as a function of the AuNPs concentration alone. This simplification can be introduced by noting that, for most hybrid polymersomes considered in this study, the scattering cross-section is at least one order of magnitude lower than the collective absorption of the nanoparticles they contain. Therefore, the former can be neglected without losing accuracy. This leads to an alternative expression for Equation 2 as (see SI):

$$\Delta T(n_{\text{AuNP}}) = \frac{P}{\beta} \left(1 - e^{\sigma_{\text{abs}} l \, n_{\text{AuNP}}}\right) \qquad (3)$$

This generalization eliminates the dependence on the polymersome morphology and allows us to predict the experimental trend emerging from all our data points analytically (Figure 3d). Once again, the points are all contained in the confidence interval defined through the limiting absorption cross-sections (Figure 3c). Therefore, the global photothermal response of hybrid polymersome suspensions with polymersome diameter up to 100 nm can be predicted with reasonable accuracy only by knowing the number density of AuNPs. Of course, this simple model only works under the assumption that the AuNPs in all the formulations have equal or comparable sizes. Such condition, as discussed in Section 2.1, can be easily guaranteed by the templating growth mechanism offered by the in situ synthesis reaction.

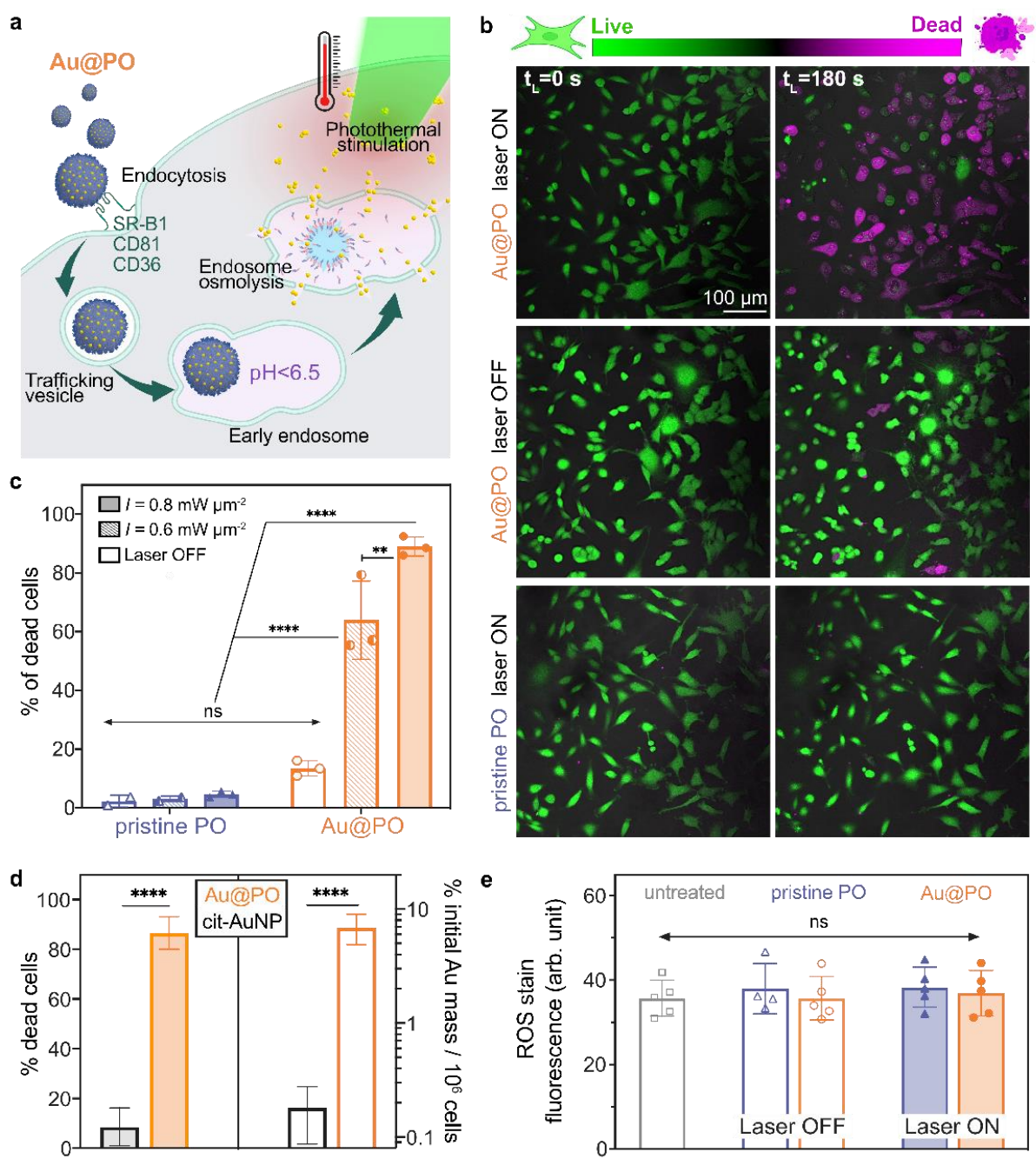

**Figure 4. In vitro photothermal therapy with hybrid polymersomes.** a) Schematic mechanism of the cellular delivery by hybrid polymersomes (Au@PO) as nano-photothermal agents. Left to right: Au@PO interact with the SR-B1, CD36 and CD81 receptors expressed on the surface of cells, promoting endocytosis. The Au@PO is transported via a trafficking vesicle to the endosome, where protonation of the pH-sensitive PDPA membrane drives the degradation of Au@PO and AuNP assemblies are released inside the cell due to osmotic shock. Laser illumination stimulates a localized temperature increase that induces cell death.

b) Fluorescence microscopy composite images of live/dead assays showing the death of T98G glioblastoma cells by Au@PO after three minutes of cumulative exposure to a 514 nm laser at 0.8 mW µm$^{-2}$ (top row). Cells treated with Au@PO without laser exposure (middle row), and with pristine polymersomes (PO) under laser exposure (bottom row) show a significantly reduced and null death rate, respectively. All cells were stained with calcein AM (live, green) and BOBO-3™ iodide (dead, magenta). c) Mortality of T98G cells treated with Au@PO or pristine PO and cumulatively exposed for three minutes to a 514 nm scanning laser at $I = 0.8$ mW µm$^{-2}$ (solid columns), $I = 0.6$ mW µm$^{-2}$ (striped columns) and without laser exposure (clear bars) as determined by live/dead assays. d) The enhanced photothermal efficacy of Au@POs originates from their cell internalization. Left: The laser-induced mortality of T98G cells treated with Au@POs is significantly higher than in cells treated with a 2x higher dose of citrate-capped gold nanoparticles (cit-AuNPs). Error bars are the standard deviations of three (cit-AuNP) and seven (Au@PO) independent experiments. Right: Quantification of the Au mass uptake by T98G cells after one hour of incubation with Au@POs shows an almost 40-fold higher uptake compared to cit-AuNPs. Error bars are the standard deviations of four (cit-AuNP) and eight (Au@PO) independent experiments. e) Reactive oxygen species (ROS) evolution measured by fluorescent assay on T98G cells treated with both Au@PO and pristine PO showed no significant difference compared to untreated cells, whether the laser was ON or OFF. In c and e, data points represent independent experiments; columns represent means, and error bars are standard deviations. Statistical significance was evaluated by two-way ANOVA for c and e, and by Student's t-test for d (ns p>0.05, ** p≤0.01, **** p≤0.0001).

2.4. Application to photothermal therapy: an in vitro validation.

We have so far demonstrated that hybrid polymersomes produced by the in situ reaction possess excellent morphological uniformity and stability; moreover, the proposed system has the ability to produce plasmonic heat boding well for photothermal therapy (PTT) applications. Based on our previous work,[35,36] we expect that PMPC-PDPA polymersomes will facilitate the cytosolic delivery of AuNPs within cells promptly and more efficiently than pristine AuNPs (**Figure 4a**). The phosphorylcholine moiety of the PMPC block in the brush is known to selectively bind to a subset of receptors overexpressed on many different cancer cell lines: SR-B1, CD36 and CD81.[37,44–47] Such affinity drives the enhanced endocytosis of the hybrid polymersomes which are transported through a trafficking vesicle to the endosome where the pH is below the PDPA pKa. Inside the endosome, the PDPA block gets protonated and drives the disassembly of the hybrid polymersome. We reproduced the endosomal conditions ($T$ = 37°C, pH 6) in the hybrid polymersome suspensions and observed by DLS and TEM that polymersomes disassemble and release AuNPs, which appear to aggregate and lose colloidal stability (**Figure S6**).

We tested two hybrid polymersome formulations as PTT agents on T98G human glioblastoma cells (Methods). These cells express the two main PMPC target receptors, i.e., SR-B1 and CD36, as demonstrated by western blot analysis on T98G cell lysates (**Figure S7a**). We incubated T98G cells with a dose of hybrid polymersomes corresponding to a gold concentration [Au] = 0.4 ppm, at which > 90% of the cells were viable after 90 minutes of incubation (**Figure S7b**). We applied a 514 nm laser to excite the LSPR and thus activate PTT using a scanning confocal microscope providing a cumulative exposure time of only three minutes over the 90 minutes of treatment. We monitored the real-time cell mortality via a live/dead fluorescence imaging assay (**Video S2**) using two imaging lasers, 488 and 562 nm, for the excitation of the fluorophores labeling the live and dead cells, respectively. The intensity of both the imaging lasers was set to a 50-fold lower value than the LSPR excitation laser.

For reference, we conducted negative control experiments monitoring the real-time mortality in the absence of the plasmon excitation laser ('laser OFF' condition, **Video S3**), and on cells treated with an equivalent dose of pristine polymersomes both with (**Video S4**) and without laser exposure. Some representative microscopy images at the beginning and at the end of the laser exposure are presented in the first and second columns of **Figure 4b**, respectively. In the cultures treated with hybrid polymersomes, the number of dead cells detected is remarkably higher after laser irradiation (top row) than in the laser OFF conditions (middle row). Control cells treated with pristine polymersomes (bottom row) present only very sporadic and isolated dead cells, indicating that the laser irradiation alone was not noticeably phototoxic within the experimental timeframe. The bar chart in **Figure 4c** provides a quantitative and comprehensive analysis of the mortality at the end of the in vitro PTT treatment, as quantified by cell counting on live/dead stained T98G. In cells treated with hybrid polymersomes, the cell death rate depends on the laser intensity to which they were exposed, with mortalities that are maximum at 89% for $I$ = 0.8 mW µm$^{-2}$, decreased to 67% for $I$ = 0.6 mW µm$^{-2}$ and dropped to only 13% when the laser stimulus was removed. As the differences between these values are statistically significant, we can infer that the mortality rate depends on the laser intensity, as does the temperature increase in our hybrid polymersomes (Figure S5).

The residual mortality recorded in the laser OFF case is ascribable to the synergetic effects of two causes: 1) the limited intrinsic toxicity measured in the viability assay (Figure S7b); and

2) the stray excitation of the plasmon resonance by the imaging lasers, which both fall within the broad LSPR band of hybrid polymersomes. In contrast, the death rate in the cells treated with pristine polymersomes was minimal at 2-4% mortality, and changes in the laser intensity did not produce any statistically significant difference in the results, thus confirming the absence of non-plasmonic phototoxicity.

The fast cell death rates detected for hybrid polymersomes, with almost 90% mortality measured only 1.5 h after the start of the PTT treatment, suggest that our systems might kill cancer cells via a non-apoptotic route. Further investigation will be needed to assign a specific pathway and determine whether an immune response is activated.

After demonstrating the applicability of our hybrid polymersomes to in vitro PTT, we experimentally validated their efficacy over non-targeted plasmonic systems. According to our proposed mechanism, the observed levels of laser-stimulated cytotoxicity are enabled by the accumulation of gold assemblies via receptor-mediated intracellular delivery. To experimentally support this argument, we repeated the in vitro PTT treatment using non-targeting citrate-capped gold nanoparticles (cit-AuNPs) at a double gold concentration than that of hybrid polymersomes, [Au] = 0.8 ppm. As shown on the left-hand side of Figure 4d, despite the higher concentration, cit-AuNPs induced an almost 10 times lower mortality than hybrid polymersomes, with only 8.6% of the illuminated cells dying during the exposure. Since, in Section 2.3, we demonstrated that the heat generation primarily depends on the AuNP concentration, such difference in photothermal effects must derive from a different final accumulation of gold caused by the interactions of the two delivery systems with the cells. To further investigate these interactions, we quantified the cellular uptake of Au in T98G cells by MP-AES after 60 minutes of incubation with either the hybrid polymersome or the cit-AuNPs suspensions (Methods). As we can appreciate on the right-hand side of Figure 4d, the cellular uptake of Au from receptor-binding hybrid polymersomes was 38 times more efficient than from cit-AuNPs. It is well known that when AuNPs are internalized within the heterogeneous cell environment, they tend to cluster.[48–50] The resulting increase in the local AuNP concentration is expected to locally maximize plasmonic heat generation and, in turn, the killing of cancer cells.[51]

We corroborate the photothermal origin of the laser-induced cell death excluding the generation of reactive oxygen species (ROS).[52] To monitor the incidence of photochemical degradation, T98G cells were first subjected to the PTT therapy treatments in all the previously tested conditions. Consequently, the treated cells were labeled with a proprietary dye that generates far-red fluorescence upon reaction with ROS (Methods). Fluorimetry measurements were collected at the exposed areas at the center of the cell-culture wells. The fluorescence intensity data presented in Figure 4e indicate that no significant difference in the concentration ROS could be detected by this assay between the untreated cells and any of the cells subjected to the different treatment conditions.

## 3. Conclusions

In conclusion, the development of biocompatible thermoplasmonic hybrid polymersomes that facilitate efficient cellular uptake and controlled loading of gold nanoparticles (AuNPs) holds great promise for targeted photothermal therapy. We have shown that the in situ synthesis of gold within polymersome membranes allows for excellent control over the loading of AuNPs with a final size templated by the membrane. The resulting hybrid polymersomes maintain colloid stability for over a year. We optimized the formulation protocol to maximize the thermoplasmonic performance of the hybrid systems, reaching temperature increments for the dispersions well over 10K. The proposed theoretical model is in good agreement with our experimental measurements, validating our predictions' accuracy and contributing to understanding the underlying photothermal mechanism. The successful combination of experimental and computational procedures de facto provides a robust foundation for the future optimization and design of targeted photothermal therapy agents based on hybrid polymersomes. Finally, our "proof-of-concept" in vitro experiments on T98G human glioblastoma cells demonstrate the remarkable photothermal cytotoxicity of hybrid polymersomes, with up to 90% cell mortality achieved within a short time frame.
These findings highlight the potential of the proposed hybrid polymersomes as effective and selective agents for future cancer therapeutic applications.

## 4. Methods

*Self-assembly of PMPC-PDPA polymersomes:* PMPC$_{20}$–PDPA$_{80}$ copolymer was synthesized by atom transfer radical polymerization (ATRP) according to a previously published protocol.[24,53] PMPC–PDPA polymersomes were fabricated by the bottom-up solvent switch method. In a typical experiment 20 mg of PMPC$_{20}$–PDPA$_{80}$ were dissolved in THF/MeOH (1:3). After complete dissolution, 2.3 mL of aqueous buffer (water, phosphate buffer 0.1M or phosphate buffer saline (PBS) 0.1M) were injected by an automatic syringe pump at 2.0 µL/min into the polymer solution stirring at 40°C. Upon completion, the reaction was quenched by adding 2.7 mL of the aqueous buffer solution at once and dialyzed against PBS 0.1 M overnight with a 3.5kDa molecular cut-off membrane (Repligen, USA). The polymersome suspensions were then purified by size exclusion chromatography (SEC) through a Sheparose 4B column and re-concentrated by tangential flow filtration through a hollow fiber module with 100 kDa MWCO (Repligen, USA).

*Synthesis of hybrid polymersomes:* Adequate volumes of a 50 mM HAuCl$_4$:3H$_2$O stock solution to reach the desired Au/DPA mol% were added dropwise to a PMPC-PDPA polymersome suspension under vigorous stirring in an ice bath. The sample was then kept stirring under pH monitoring to allow the partial protonation of the PDPA blocks and the diffusion of the AuCl$_4^-$ ions to the membrane. After 30 min, an equivalent volume of freshly prepared ice-cold 50 mM NaBH$_4$ solution was added to the sample at once. Samples were immediately transferred to a 3.5 kDa MWCO dialysis membrane and dialyzed against PBS 0.1 M to remove unreacted species. To eliminate AuNPs that may nucleate in the solvent after membrane saturation, the Au-polymersome samples were purified by repeated cycles of tangential flow filtration through a hollow fiber module with 0.05 µm cut-off (Repligen, USA) until a filtrate volume of 40x the sample was collected.

*Morphological characterization of pristine and hybrid polymersomes:* The size distribution of pristine and hybrid polymersomes was determined by dynamic light scattering (DLS) on a Malvern Zetasizer Nano ZS (Malvern Ltd., UK) instrument equipped with a 120 mW He-Ne laser (630 nm) at a controlled temperature of 25 °C in a 173° detector configuration. Samples were diluted to 0.1 mg/mL with PBS 0.1 M before the measurement. The morphology of the formulations was determined by transmission electron microscopy (TEM). TEM specimens were prepared by drop-casting 5 µL of sample onto a carbon-coated Cu grid, previously glow-discharged for 45 s. After 1 min, the excess sample was drained from the grid by blotting with filter paper. Positive staining of the polymer was applied by immersing the grid for 3 s in a drop of 0.5 wt% phosphotungstic acid (PTA) aqueous solution. The excess staining solution was drained again, and the grid was further dried under vacuum for 1 min. The grids were imaged at 200 kV using a JEM-2100 TEM (JEOL, JP) equipped with a Gatan Orius SC-200 camera.

*Concentration measurements:* The final polymer concentration was determined by reverse phase high performance liquid chromatography (RP-HPLC) using a Jupiter C18 column (Phenomenex, USA). Before the analysis an aliquot of each sample was dissolved into a 24:1 PBS pH 1/NaCl 1M acid mixture to solubilize the polymer and centrifuged three times at 20000 RCF for 1h at room temperature to precipitate the AuNPs. The supernatant was passed through the HPLC column using a mixture of methanol/water in a linear concentration gradient as a mobile phase to allow isolation of the polymer component. The polymer concentration was quantified from the UV absorption at 220 nm.
The Au concentration was measured by microwave assisted plasma atomic emission spectroscopy (MP-AES). An aliquot of each sample was dissolved in a 2x volume of aqua regia (3:1 HCl/HNO$_3$), sonicated for 45 min and left overnight to allow for the complete solvation of the Au ions. The solutions were then diluted 10:1 with milliQ water and analyzed with an Agilent 4100 MP-AES instrument. Au was detected by the emission lines at 242.795 nm and 267.595 nm and the concentration was determined from the calibration of the instrument with an ICP grade Au calibration standard (5190, Agilent, USA).

*UV-vis absorption spectroscopy:* UV-vis extinction spectra of pristine and hybrid polymersomes were collected with a Shimadzu UV-2700 UV-vis spectrophotometer using quartz cells with a light path length $l$ =10 mm and PBS 0.1 M as a reference. The extinction value was indirectly calculated as $E = -\log_{10} \Phi_t/\Phi_0$ where $\Phi_t$ and $\Phi_0$ are the transmitted and incident photon fluxes, respectively. Absorption spectra of hybrid polymersomes were calculated by subtraction of the extinction spectra of pristine polymersomes at the same concentration, under the assumption of negligible absorption of polymersomes and negligible scattering of AuNPs in the visible range.

*Temperature measurements:* A 300 μL volume of aqueous suspension of pristine or hybrid polymersomes was transferred to a 10 mm path length quartz cuvette and exposed to a solid state continuum 532 nm laser beam (Gem 532, Laser Quantum, UK) focused through a plano-convex lens with foal length $f = 50$ mm (LA1131-A-ML, Thorlabs, DE). Where not specified the nominal laser power was set to 200 mW, corresponding to an incident power on the cuvette of 180 mW. Meanwhile, the evolution of temperature was monitored through an in-house resistance temperature detector (RTD) composed of a 4-wire PT-100 probe and a MAX31865 Analog-to-Digital (RS Components, UK) mounted onto an Arduino Uno board (Arduino CC, Italy), offering a sensitivity of ± 0.03 K. The data acquisition Arduino library 'Adafruit_MAX31865' was developed under a BSD license by Limor Fried/Ladyada at Adafruit Industries (US). For selected samples the spatial evolution of the temperature upon irradiation in the cuvette was imaged through a FLIR A6703 IR camera with a f/4.0 aperture and 60Hz frame rate.

*T98G cell culture:* The T98G human glioblastoma cells were kindly provided by Prof. Vivaldo Moura Neto (IECPN, Brazil). This cell line originated from a white, 61 years-old male and presents a fibroblast-like cell phenotype. Cells were cultured in Dulbecco's modified Eagle Medium-GlutaMAX with glucose and sodium pyruvate, supplemented with 10% fetal bovine serum (FBS) and 1% penicillin/streptomycin mixture at 37°C in a 5% $CO_2$ atmosphere. For subcultures, after reaching confluency, the cells were detached from the flasks, using 0.25% (w/v) trypsin/EDTA for 5 min at 37°C and centrifuged at 200 RCF for 5 minutes. Different cellular densities were used according to each experiment.

*MTT viability assay:* T98 cells were cultured into 96-well plates at $1 \times 10^4$ cells per well, overnight at 37°C in a 5% $CO_2$ atmosphere. Then, cells were incubated with serial dilutions of stock concentrations of pristine polymersomes (POs) and hybrid polymersomes (Au@POs) (max. polymer concentration $c = 0.1$ mg/mL) for 1.5 h. Thereafter, cell viability was assessed by the MTT reduction colorimetric assay by adding 50 µL of MTT to each well (0.5 mg/mL). After two hours of MTT incubation, the production of the purple formazan crystals was observed, which were then dissolved by adding 100 µL of DMSO. Absorbances were measured at 570 nm in a plate reader.

*Western blotting:* To generate cell lysates, $10^6$ T98G cells were washed with PBS and centrifuged at 400 RCF, twice. Cell pellets were then lysed with 100 µL of RIPA buffer (Sigma-Aldrich, DE) including protease inhibitors (1:100, Sigma-Aldrich, DE) for 30 min at 4°C. Cell lysates were further centrifuged at 17000 RCF, 4 °C for 20 min, after which supernatants were collected and protein content quantified by Bradford assay following the manufacturer's protocol (Protein Assay Dye Reagent, Bio Rad).

For western blot analysis, proteins were denatured with 2x Laemmli buffer (Bio-Rad) and β-mercaptoethanol (Bio-Rad) at 95 °C for 5 min prior to sample addition into 12% Bis-Tris acrylamide gel containing 10% sodium dodecyl sulfate (SDS). A total of 20 µg of protein (40 µL maximum per well) was added to gels, on which electrophoresis was run for 30 mins at 80V, followed by 1.5 hours at 120V using a Bio-Rad Power Pac source system in 1X Tris-Glycine running buffer. After electrophoresis, proteins were transferred into polyvinylidene difluoride (PVDF) membranes (Bio-Rad) using a wet transfer system in Tris-Glycine buffer with 20% methanol at 50 V for 1 hour, followed by 100 V for 1 hour at 4°C. Subsequently, the PVDF membrane was blocked with 5% non-fat dried milk powder (PanReac, AppliChem) in Tris-Buffer Saline (TBS) for 1 hour at room temperature. Membranes were then incubated with diluted primary antibodies to an adequate working concentration in a solution of 1% milk in TBS+0.1%Tween (TBS-T) at 4°C for detecting the specific scavenger receptors: anti-SR-B1 from rabbit (Novus Biologicals, NB400-131SS, dilution 1:1000) and anti-CD36 from rabbit (Novus Biological NB400-144SS, dilution 1:500). Membranes were further probed for Glyceraldehyde 3-phosphate dehydrogenase using a-GAPDH from mouse (Proteintech, 60004-1-Ig, dilution 1:10000) as a loading control. Next, the membranes were washed three times with TBS-T, and the corresponding diluted secondary antibodies were incubated for 1 hour at room temperature and washed again three times with TBS-T. Membranes were imaged on a Licor CXF system (Odyssey) and analyzed in the ImageJ software.

*Au cellular uptake of T98G cells:* Initially, $5 \times 10^5$ T98G cells/well were plated in tissue culture-treated 6 well plates and grown for 24h. In each well, the cells were incubated with 1 mL of the appropriate culture medium supplemented with fetal bovine serum (FBS) and containing ≈$4.5 \times 10^{11}$ Au- polymersomes (total Au mass: 6.4 µg) for 1 h at 37°C and 5% $CO_2$. In control experiments, cells were treated with comparable doses of pristine PMPC-PDPA polymersomes or commercial 1.8 nm citrate AuNPs dispersed in the FBS-supplemented medium. After incubation, the cells were washed 3x with PBS to eliminate the excess nanoparticles and detached by incubation with 1 mL trypsin/EDTA for 5 min. After centrifugation at 200 RCF for 5 minutes the pellets were redispersed in 1 mL of fresh culture medium for counting with a TC20 automated cell counter (Bio-Rad Laboratories Inc, US). After reprecipitation, the obtained pellets were treated with 150 µL of aqua regia, sonicated for 45 minutes at room temperature, and left overnight to allow for the complete digestion of the organic matrix and solvation of the Au ions. The solutions were then diluted 10:1 with milliQ water and the concentration of Au was measured with an Agilent 4100 MP-AES instrument (Au emission line at 267.595 nm) after calibration of the instrument with an ICP grade Au calibration standard (5190, Agilent, USA).

*In vitro photothermal therapy (PTT) experiments:* Initially, $1.2\times10^4$ T98G cells/well were plated on tissue-culture treated microscopy grade 8 well plates (ibidi, DE) and grown for 24 h. The Live/Dead staining kit (Invitrogen, USA) was prepared by dissolving the lyophilized 'Dead' staining agent (BOBO-3 iodide) into the 'Live' staining solution (Calcein AM). A dose of 100 nm hybrid polymersomes to give a final [Au] = 0.4 ppm was dispersed in 100 μL of Live Cell Imaging solution (Invitrogen, USA) and mixed with 100 μL of freshly prepared Live/Dead staining solution. The cell culture medium was removed from the well and the cells were incubated with the hybrid polymersome/staining solution for 15 min at 37°C, 5% $CO_2$. Control experiments were carried out using an equivalent dose of pristine polymersomes (0.01 mg/mL of polymer) or a 2x Au-dose, [Au] = 0.8 ppm, of 1.8 nm citrate-capped gold nanoparticles. Incubated cells were handled avoiding exposure to stray lights. After incubation, the cells were transferred to the environmental chamber of a Leica TCS SP8 confocal microscope where laser stimulation and real-time fluorescence imaging were carried out in alternated cycles. The laser stimulation cycle was performed by focusing the 514 nm line of an $Ar^+$ laser at 0.5 and 0.35 mW through a 20x objective (NA=0.70) on the cells and raster scanning it across the field of view with a pixel dwell time of 1.2 μs. During the imaging cycle the Live and Dead dyes were excited using the 488 nm line of the $Ar^+$ laser and the 561 nm line of a diode-pumped solid-state (DPSS) laser, respectively. Bright-field images were also collected, illuminating the cells with a 405 nm diode laser. To minimize photobleaching and photodegradation, and avoid photothermal stimulation during this cycle, the imaging lasers power was kept at $P \leq 0.01$ mW. The cycles were alternated for 90 minutes with a period of 9.25 s, leading to a total exposure time of ~ 0.7 ms/pixel and ~3 min/frame.

*Reactive oxygen species (ROS) evolution assay:* $1\times10^4$ T98G cell/well were plated in black 96-well flat-bottomed plates and grown overnight. The culture medium was replaced with fresh culture medium containing a dose of $9.2\times10^9$ Au-polymersomes. In control experiments, the cells were treated with an equivalent dose of pristine polymersomes. The cells were incubated in the environmental chamber (37°C, 5% $CO_2$) of a Leica TCS SP8 confocal microscope and the area at the center of the well was exposed to the scanning 514 nm laser line in the same illumination conditions used for the in vitro PTT experiments (0.5 mW and laser OFF conditions, 90 minutes overall treatment). After the treatment, 100 µL of Cellular ROS Deep Red staining solution (Abcam) was added to the cells and they were incubated for 45 min at 37°C, 5% $CO_2$. Untreated T98G cells were stained with the same procedure for reference. The ROS-activated fluorescence signal was detected with a Tecan Spark fluorimeter in top reading mode in the center of the well with excitation wavelength (650±5) nm and emission wavelength (675±5) nm. Each measure was collected over 30 flashes of 40 µs each. The optimal z-position was determined as the height at which the maximal fluorescence signal was detected during a preliminary z-scan.

*Optical calculations:* The extinction cross-sections of pristine and hybrid polymersomes were calculated using a Mie theory approach. The polymersome was modeled as a four-layer stratified spherical particle composed of a water core surrounded by three polymer layers: a hydrophobic PDPA layer sandwiched between two hydrated PMPC ones.
The thicknesses of the three layers were evaluated by semi-empirical scaling laws of the form $bM^v$, where $b$ is the length of the monomer, $M$ the degree of polymerization, and $v$ is the Flory exponent, $v = 0.9$ for PMPC and $v = 2/3$ for PDPA. [41]
The Mie calculations were carried out using the Scattnlay 2.0 code.[54] The hydrophobic PDPA layer was optically modeled as polymethylmethacrylate (PMMA), while the refractive index of the hydrated PMCP layers was calculated by the Bruggeman effective medium approximation[55] using dry PMPC[56] and water as boundary media. The water volume fraction $\phi_w = 0.39$ was estimated from experimental data on bound water in PMPC brushes reported by Hatakeyama et al.[57] The resulting effective refractive index is 1.43. The optical properties of ultra-small gold nanoparticles were calculated by extrapolation of the semi-empirical model by Karimi et al.[42], which gave a permittivity of $\epsilon = -5.03 + 8.51i$ at $\lambda = 532$ nm. The absorption cross-section was determined in the quasi-static approximation of the Mie theory [2] as: $\sigma_{abs} = \frac{8\pi^2\sqrt{\epsilon_m}a^3}{\lambda_0} \cdot \text{Im}\left\{\frac{\epsilon-\epsilon_m}{\epsilon+2\epsilon_m}\right\}$, where $a$ is the radius of the AuNP and $\epsilon_m$ the permittivity of PMMA. Lastly, the extinction cross-section of the hybrid polymersome was calculated as a linear combination of the scattering cross-section of an empty polymersome $\sigma_{sca}$ (the absorption was found to be negligible) and the absorption cross-sections of the $\chi_{AuNP}$ embedded AuNPs: $\sigma_{ext} = \chi_{AuNP}\sigma_{abs} + \sigma_{sca}$.

*Thermoplasmonic simulations:* The light-matter interaction of isolated hybrid polymersomes was numerically simulated by finite element methods (FEM) using COMSOL Multiphysics 5.6, a software offering integrated solid methods for the solution of partial differential problems on complex geometries. The geometry of the system under analysis was modeled as a stratified sphere (**Figure S8a**) composed of 4 elements: 1) a water core; 2-3) two hydrated PMPC layers, and 4) a dry polymer layer (hydrophobic membrane) sandwiched between the brushes, modeled as PMMA. The optical constants of the PMPC layers were calculated as described in the 'Optical calculations' section, giving 1.43. Their effective thermal conductivity $k = 0.292$ W m$^{-1}$ K$^{-1}$ was estimated by applying the Bruggeman effective medium approximation with PMMA and water as boundary media at a water volume fraction $\phi_w = 0.39$). A randomly distributed array of $\chi_{\text{AuNP}}$ gold nanoparticles with diameter $d = 2$ nm was embedded within the membrane. The optical properties of the gold nanoparticles were calculated as described in the 'Optical calculations' section. All structures were meshed with a wavelength-controlled free tetrahedral mesh having a maximum size of 0.33 nm for the AuNPs and 1 nm for the polymer layers (**Figure S8b**). A combination of scattering boundary limits (set to 12x the structure size) and perfectly matched layer (PML) domains were used as boundary conditions to avoid computational artifacts coming from parasite optical events, e.g., artificial reflections. The simulation workflow entails two distinct steps. First, the electromagnetic part of the problem was solved using the RF Module under illumination by a linearly polarized plane wave of wavelength $\lambda = 532$ nm and intensity $I = 0.128$ mW µm$^{-2}$. Second, the Heat Transfer in Solids (HTS) interface was applied to solve the heat equation considering the resistive losses calculated in the previous step as a heat source and a heat flux node (convective interface) along the outer boundaries as a heat sink.

**Supporting Information**

Supporting Information is available from the Wiley Online Library or from the author.

Acknowledgements
V.B. thanks Dr Manish Trivedi and Dr Robert Malinowski for their advice on developing the optical setups to characterize the thermoplasmonic properties. V.B. and G.V. acknowledge Guillaume Baffou for insightful discussions. V.B. acknowledges the EPSRC via a DTA for sponsoring his PhD. J. G-C. thanks the Ministry of Science and Innovation of Spain for his FPI grant and P.A. acknowledges funding for a Ramon y Cajal Fellowship [Grant No. RYC-2016-20831]. D.M thanks the EPSRC for the SomaNautix Healthcare Partnership

EP/R024723/1 and CRUK Edinburgh-UCL Brain Tumour Centre of Excellence Award (177884) . G.V. gratefully acknowledges the Engineering and Physical Sciences Research Council for supporting this work [grant numbers EP/W005875/1, EP/R513143/1] and for providing A.T. with a research scholarship [grant number EP/R513143/1]. A.D.-C. acknowledges the Royal Society (Newton International Fellowship scheme 2017-NF171487) and EU H2020 MSCA–IF-792957 SPeNTa-Brain. G.B. thanks the ERC ChessTaG (769798), and the Ministry of Science and Innovation of Spain (Proyectos I+D+I PID2020-119914RB-I00), for sponsoring part of this work.

Received: ((will be filled in by the editorial staff))
Revised: ((will be filled in by the editorial staff))
Published online: ((will be filled in by the editorial staff))

**TOC**

**Designing thermoplasmonic polymersomes for photothermal therapy**

*Valentino Barbieri[1,2,3], Javier González-Colsa[4], Diana Matias[1,2,5], Aroa Duro-Castano[1], Anshu Thapa[1], Lorena Ruiz-Perez[1,2,3,6], Pablo Albella[4], Giorgio Volpe[1,*] and Giuseppe Battaglia[1,2,3,7]\**

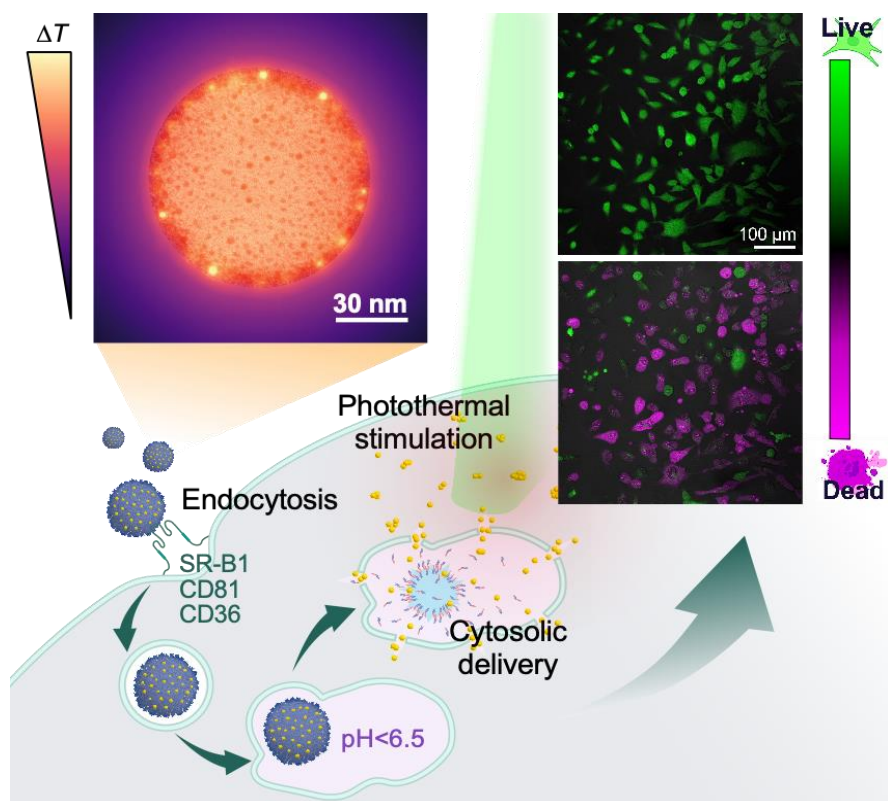

We developed polymer vesicles, known as polymersomes, with gold nanoparticles embedded in their membrane. Upon laser illumination, these hybrid polymersomes show excellent thermoplasmonic properties, which we characterize experimentally and describe analytically. These properties, together with the enhanced intracellular delivery capabilities of polymersomes, enable in vitro photothermal therapy of cancer cells.

# Supporting Information

**Designing thermoplasmonic polymersomes for photothermal therapy**

*Valentino Barbieri[1,2,3], Javier González-Colsa[4], Diana Matias[1,2,5], Aroa Duro-Castano[1], Anshu Thapa[1], Lorena Ruiz-Perez[1,2,3,6], Pablo Albella[4], Giorgio Volpe[1,\* and Giuseppe Battaglia[1,2,3,7]\**

[1]Department of Chemistry, 20 Gordon Street, University College London, WC1H 0AJ London, United Kingdom.

[2]Molecular Bionics Group, Institute for Bioengineering of Catalunya (IBEC), The Barcelona Institute of Science and Technology (BIST) Barcelona, Spain.

[3]Biomedical Research Networking Center in Bioengineering, Biomaterials, and Nanomedicine (CIBER-BBN), Barcelona, Spain.

[4]Group of Optics, Department of Applied Physics, University of Cantabria, 39005, Santander, Spain.

[5]Instituto de Medicina Molecular João Lobo Antunes (iMM), Lisbon, Portugal.

[6]Department of Applied Physics, University of Barcelona, Barcelona, Spain.

[7]Catalan Institution for Research and Advanced Studies (ICREA), Barcelona, Spain.

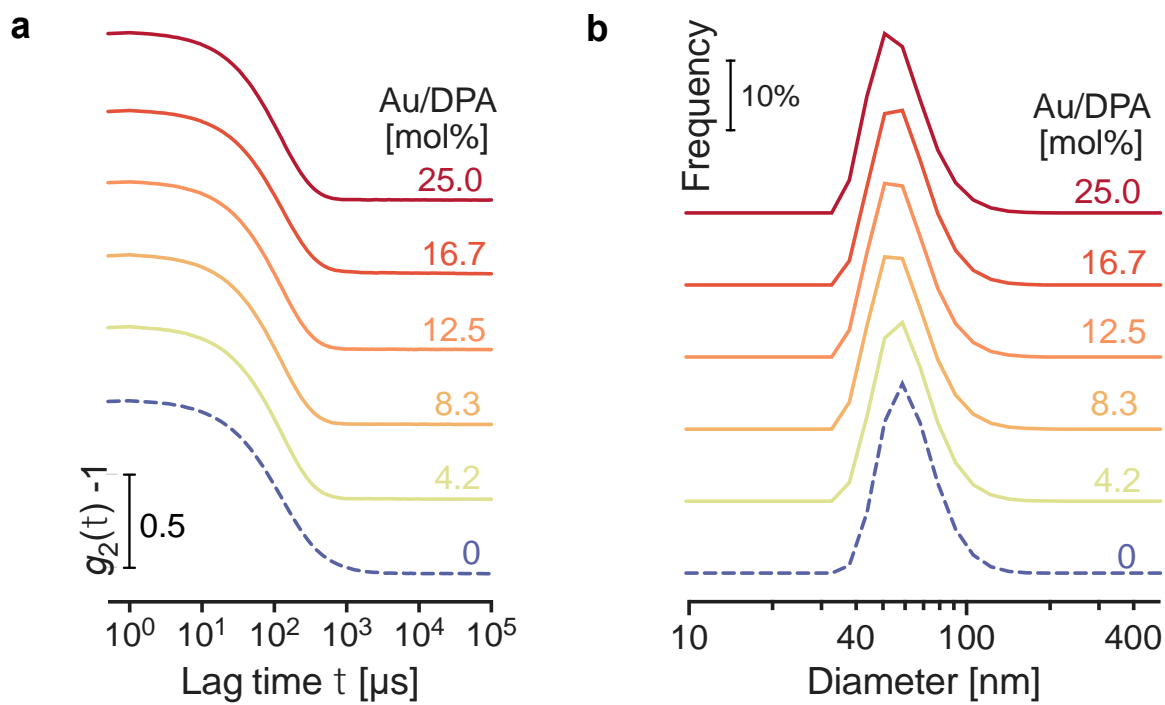

**Figure S1. Size distributions of hybrid polymersome by dynamic light scattering (DLS).** a) Representative DLS autocorrelation functions $g_2(\tau) - 1$ of hybrid polymersomes suspensions synthesized at increasing Au/DPA molar percentages, compared to the pristine polymersomes. All the samples show a single exponential decay at comparable times. b) Number-averaged DLS size distributions (derived from the distribution analysis of the curves in a) do not vary significantly after the synthesis of AuNPs in their membrane up to 25 mol% Au/DPA. Shaded areas represent the standard deviation from triplicates.

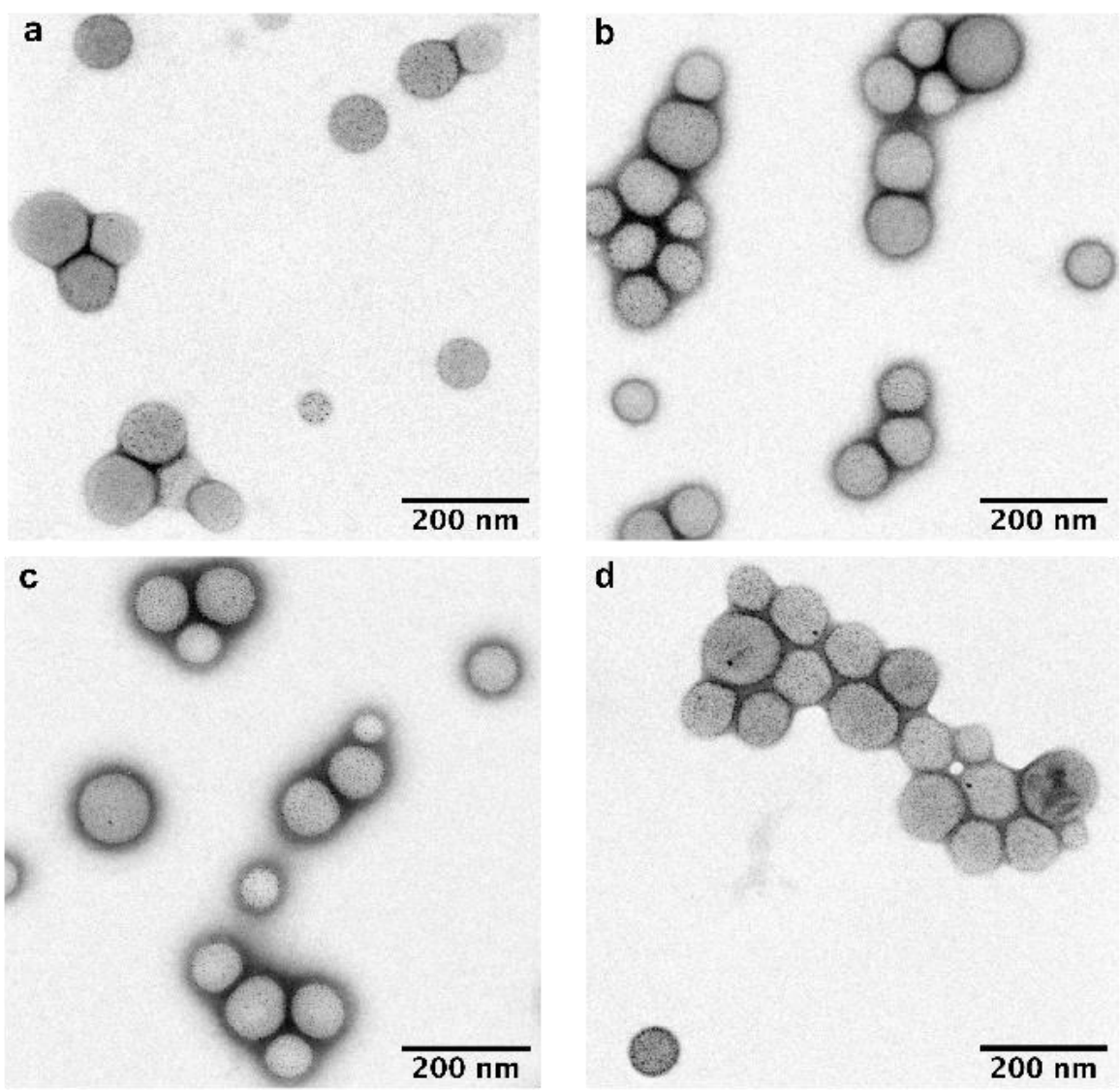

**Figure S2. Additional TEM images.** Low magnification transmission electron microscopy (TEM) images of hybrid polymersomes synthesized at Au/DPA ratios of a) 3.1 mol%, b) 6.2 mol%, c) 12.5 mol% and d) 25 mol%.

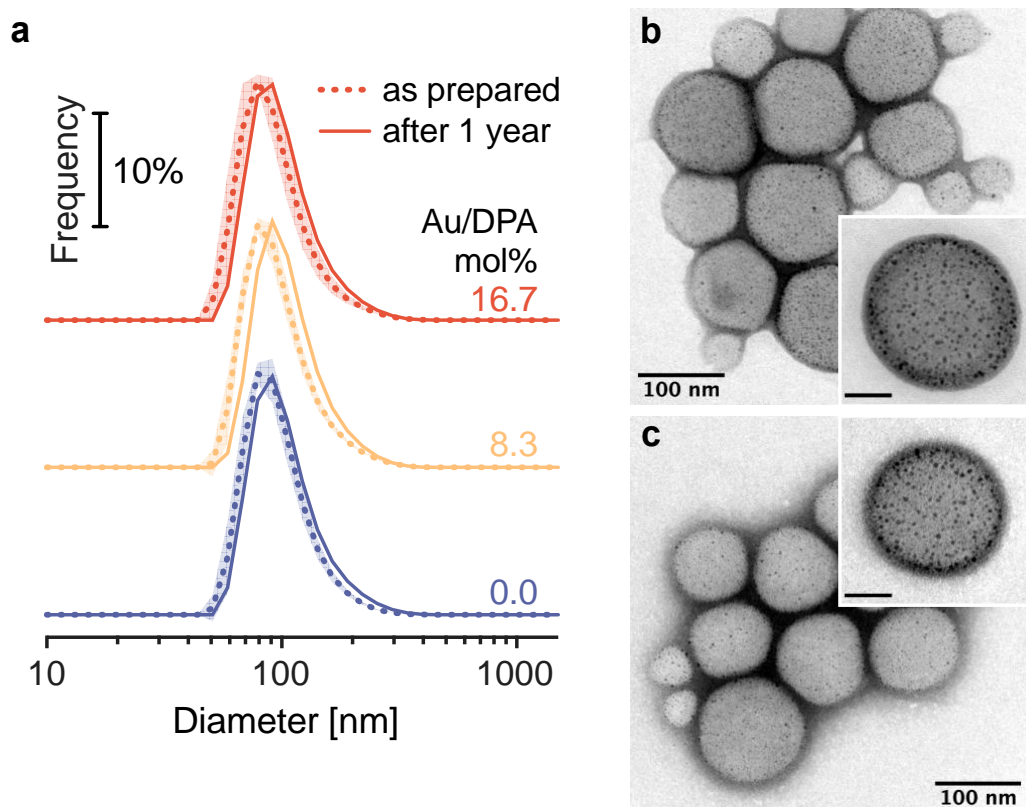

**Figure S3. Morphological characterization of the long-term stability of hybrid polymersomes.** a) Number-averaged DLS size distributions of pristine and hybrid polymersomes suspensions measured after preparation (dotted lines) and after 1 year of storage at 4°C (solid lines). Shaded areas represent the standard deviation from triplicates. b-c) TEM images of hybrid polymersomes b) after preparation and c) after 1 year of storage at 4°C. The insets show higher magnification images of isolated hybrid polymersomes (scale bars: 30 nm).

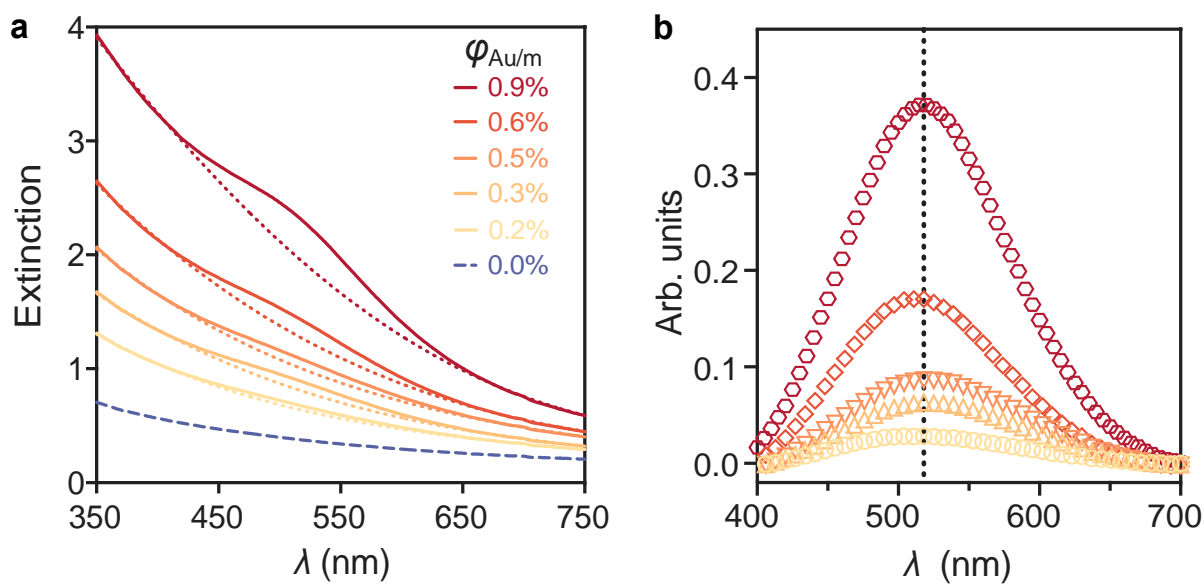

**Figure S4. UV-vis extinction spectra of hybrid polymersomes**. a) Representative UV-vis extinction spectra of hybrid polymersomes with different volume ratios of Au in the membrane $\varphi_{Au/m}$ (solid lines). Dotted lines represent best polynomial fits of the spectra outside the plasmon resonance bands. b) The localized surface plasmon resonance (LSPR) wavelength is estimated by finding the maxima of the peaks obtained by subtracting the polynomial fits from the spectra in a.

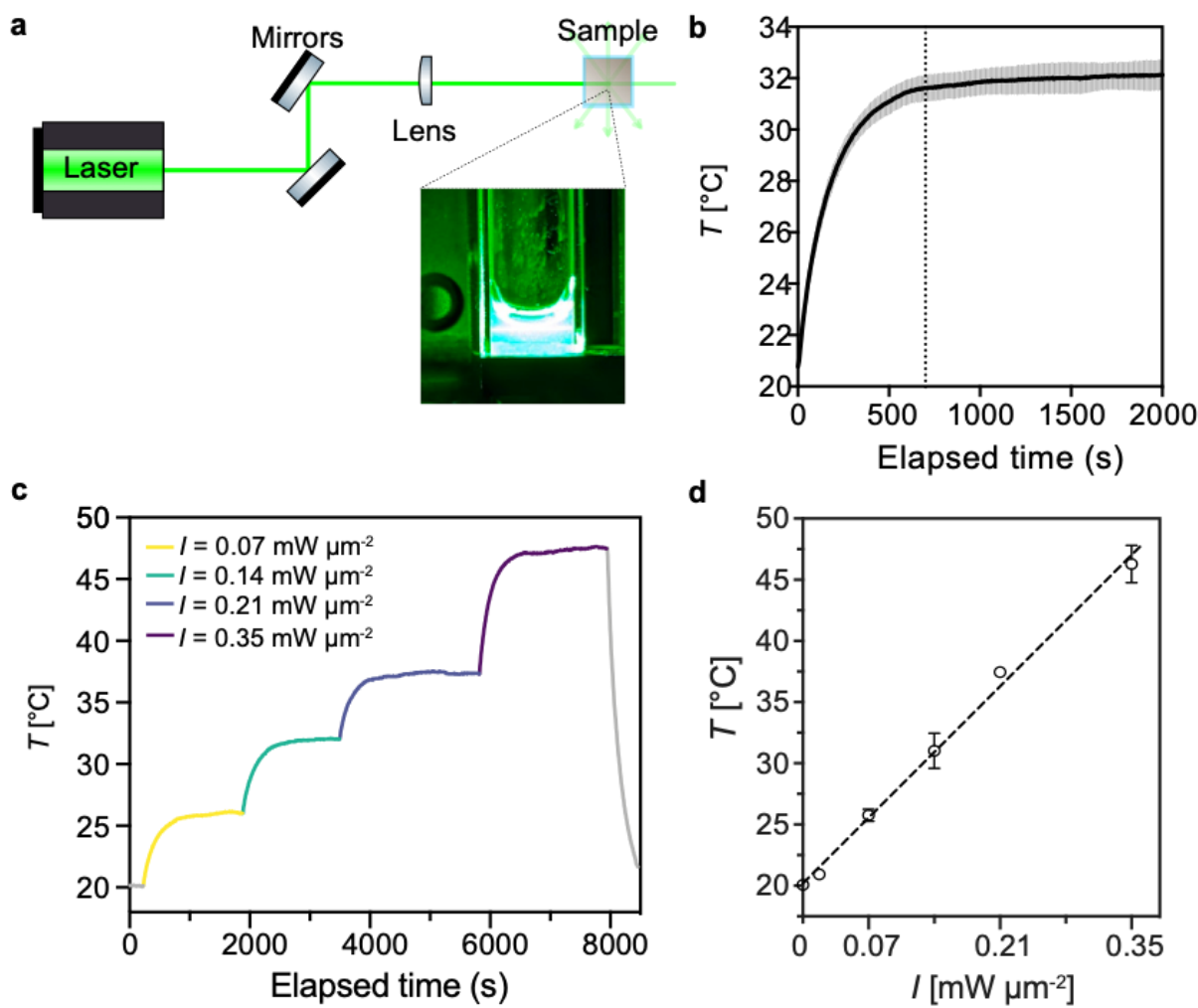

**Figure S5. Characterization of hybrid polymersome thermal properties.** a) Scheme of the optical setup used for the thermoplasmonic characterization of hybrid polymersomes. Inset: photograph of a sample under laser irradiation. b) Typical time evolution of the temperature in a hybrid polymersome suspension irradiated at 532 nm with an intensity $I = 0.128$ mW µm$^{-2}$, showing that the steady state is reached after ~700s (vertical dotted line). The shaded area represent the standard deviation from triplicates. c) Thermal response of a hybrid polymersome suspension irradiated at 532 nm with a stepwise increase of the incident laser intensity. d) The equilibrium temperature obtained from the previous plot depends linearly on the laser intensity.

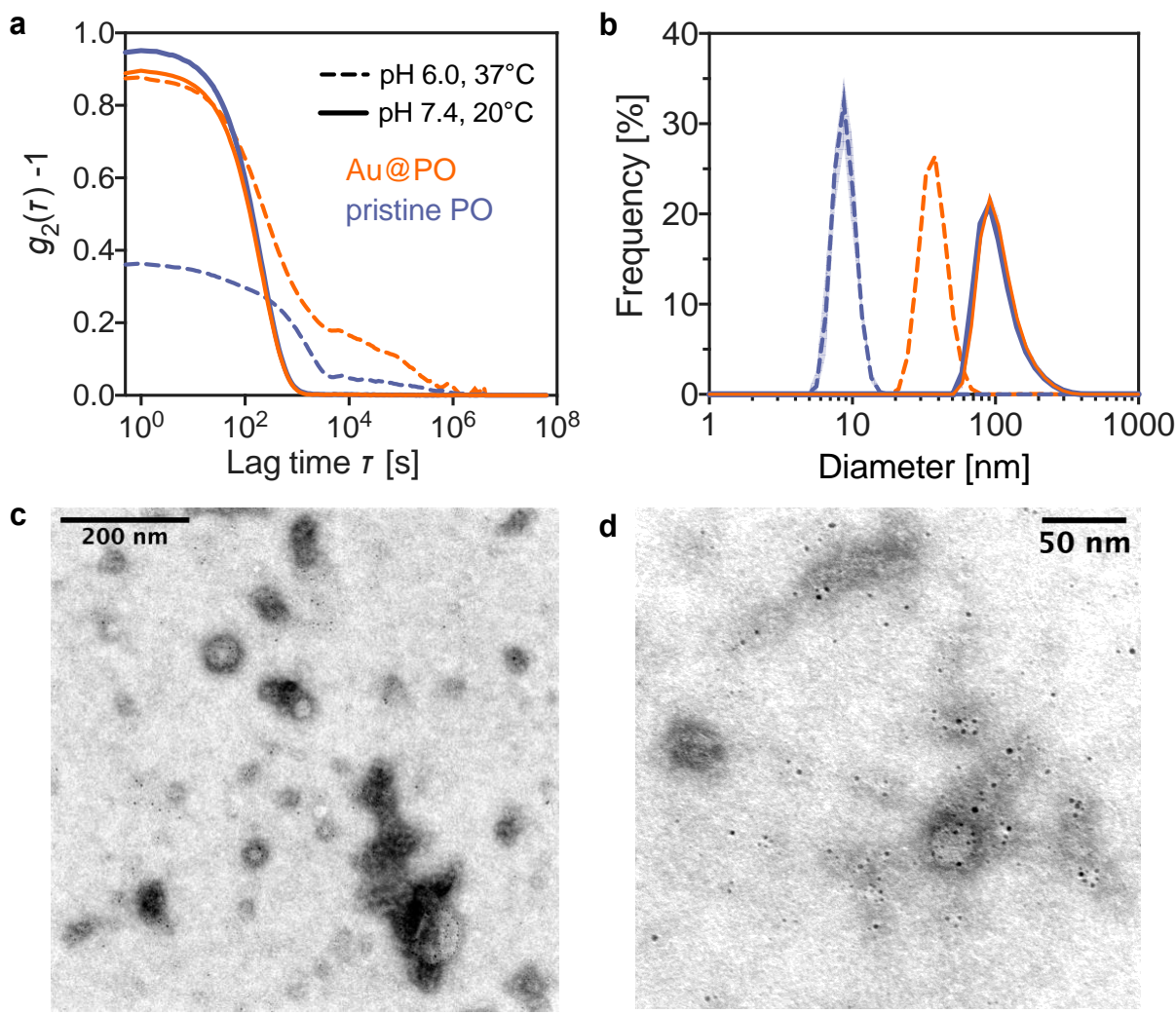

**Figure S6. Degradation study of hybrid polymersomes at the endosomal pH and temperature** a) DLS autocorrelation functions $g_2(\tau) - 1$ of pristine (blue) and hybrid polymersomes (orange) at pH 7.4 and 20 °C (solid line) and after 1.5 h at pH 6.0 and 37 °C (dashed lines). b) DLS number-averaged distribution calculated by distribution analysis of the correlation functions in a. In a and b, the shaded area corresponds to one standard deviation of three repeated measures. c-d) TEM images of hybrid polymersome sample after 1.5h at pH 6 and 37°C, revealing the presence of small polymer-stabilized gold clusters and isolated AuNPs.

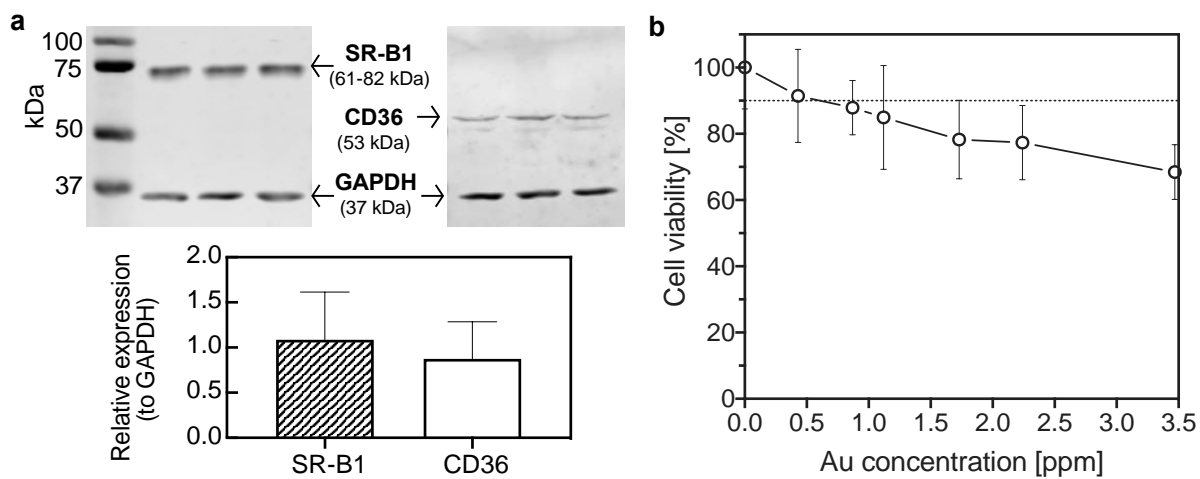

**Figure S7. Evaluation of receptor expression and viability of T98G** a) Top: Representative Western blot membranes of T98G cells lysates revealing the expression of SR-B1 (left) and CD36 (right) receptors, compared to the GAPDH as a loading control. Bottom: Quantification of the relative expression of the two receptors compared to GAPDH. b) Viability of T98G cells evaluated by colorimetric MTT assay after 1.5 h of incubation with hybrid polymersomes. The dotted horizontal line represents the toxicity threshold of 90% viability.

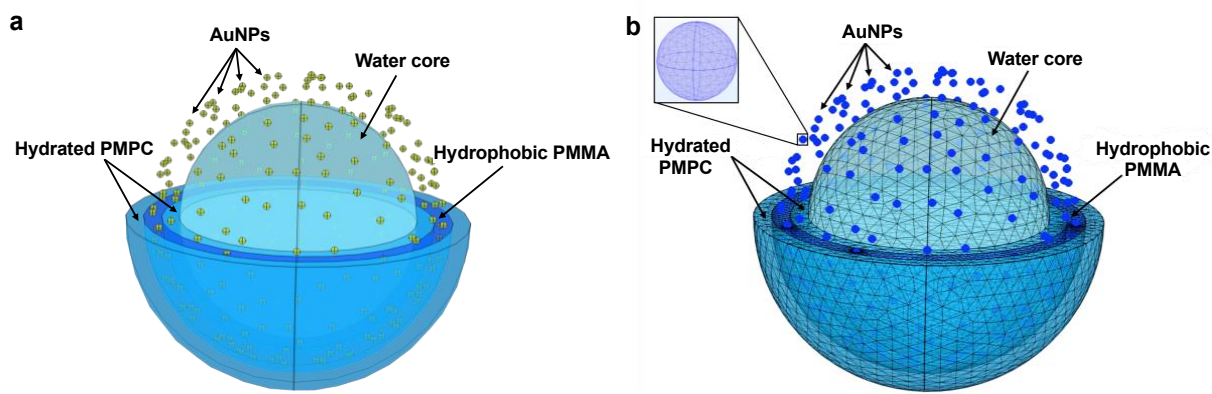

**Figure S8. Simulated geometry.** a) Geometry of a single representative hybrid polymersome built in Comsol Multiphysics fort the finite elements simulations. b) Meshing of the hybrid vesicle represented in a.

**Au-loading model**

The number density of polymersomes $n$ and the total membrane volume $V_m$ were determined combining concentration data and DLS size distributions. The number density of polymersomes having radius $R_i$ in the discrete size distribution $f_i = f(R_i)$ is:

$$n_i = \frac{N_i}{N_{\text{agg},i}} \qquad (S1)$$

with $N_i$ being the total number of polymer molecules in the bin $i$ and $N_{\text{agg},i}$ is the aggregation number, i.e., the number of polymer molecules in a polymersome calculated as:

$$N_{\text{agg},i} = \frac{pV_{\text{m},i}}{v} \qquad (S2)$$

where $v$ is the molecular volume of the PDPA block, $p$ is the packing parameter ($p \sim 1$ for vesicular systems) and $V_{\text{m},i} = \frac{4}{3}\pi(R_i - b)^3 - (R_i - b - m)^3$ is the membrane volume of the polymersome of radius $R_i$ having brush and membrane thicknesses $b$ and $m$, respectively. $N_i$ can be calculated as:

$$N_i = cN_A \cdot \frac{f_i N_{\text{agg},i}}{\sum_i f_i N_{\text{agg},i}} \qquad (S3)$$

where $c$ is the molar concentration of polymer measured by RP-HPLC and $N_A$ the Avogadro's constant. Therefore, inserting (S3) in (S1) we obtain:

$$n_i = cN_A \cdot \frac{f_i}{\sum_i f_i N_{\text{agg},i}} \qquad (S4).$$

A weighted sum of all $n_i$ over the whole size distribution gives the total number density:

$$n = \sum_i n_i \qquad (S5).$$

At this point the total volume fraction of the PDPA membranes in the sample can be obtained as:

$$\phi_m = \sum_i n_i V_{\text{m},i} \qquad (S6).$$

The total volume fraction of gold in the sample is calculated as:

$$\phi_{Au} = \frac{c_{Au}}{10^6 \cdot \rho_{Au}} \tag{S7}$$

where $c_{Au}$ is the concentration of Au in ppm calculated by MP-AES and $\rho_{Au}$ the density of Au in g/cm³. The total number density $n_{AuNP}$ can then be obtained as:

$$n_{AuNP} = \frac{\phi_{Au}}{V_{AuNP}} \tag{S8}$$

where $V_{AuNP} = \frac{4}{3}\pi a^3$ is the volume of a particle of radius a.

Finally, the two loading indicators, i.e., the gold-to-membrane volume fraction $\varphi_{Au/m}$ and the loading density of AuNP per polymersome $\chi_{AuNP}$ are calculated as follows:

$$\varphi_{Au/m} = \frac{\phi_{Au}}{\phi_m} \qquad \chi_{AuNP} = \frac{n_{AuNP}}{n} \tag{S9}$$

**Theoretical model of the heat generation and transfer in hybrid polymersome suspensions.**

A model was elaborated to describe the evolution of the observed temperature increment $\Delta T$ upon laser illumination.

The following hypotheses were applied:

1. Gold nanoparticles are much smaller than the incident wavelength (quasi-static approximation).
2. The polymer absorption and AuNP scattering are negligible in the visible range.
3. No electromagnetic coupling occurs between AuNPs, which are considered as independent dipoles.
4. The concentration is in the validity range of the Beer-Lambert's law.
5. The incident beam has a Gaussian power distribution.
6. The concentration of the hybrid polymersomes is homogenous in the sample.

Under assumptions 1 and 2, the extinction cross section of a single hybrid polymersomes $\sigma_{ext}$ can be expressed as a linear combination of the scattering cross section of the polymer shell ($\sigma_{sca}$) and the absorption cross sections ($\sigma_{abs}$) of all the $\chi_{AuNP}$ embedded AuNPs: $\sigma_{ext} = \chi_{AuNP} \sigma_{abs} + \sigma_{sca}$.

Scattering and absorption events attenuate the intensity $I$ of the laser beam as it proceeds through the sample, so that $I = I(\vec{r}, n)$, where $\vec{r} = (x, y, z)$ is the spatial coordinate of a generic point in the illuminated volume and $n$ is the number density of hybrid polymersomes. Upon laser illumination, the LSPR also results in the generation of thermoplasmonic heating. The total generated heat flow rate is given by:[58]

$$\dot{Q}(\vec{r}, n) = \sigma_{abs} I(\vec{r}, n) N_{AuNP} \qquad (S10)$$

where $N_{AuNP}$ represents the total number of gold nanoparticles acting as absorbers/emitters, i.e., particles within the illuminated volume $V$.

Given the homogeneous random distribution across the whole suspension, we can assume that $N_{AuNP} = n_{AuNP} V$ with $n_{AuNP}$ being the number density of AuNPs, which in turn can be related to the hybrid polymersomes number density, $n$, through: $n_{AuNP} = n \chi_{AuNP}$. Therefore, in the infinitesimal volume element $dV$, an infinitesimal heat flow rate is produced:

$$d\dot{Q}(\vec{r}, n) = n \chi_{AuNP} \sigma_{abs} I(\vec{r}, n) dV \qquad (S11).$$

In the absence of the sample, the spatial dependence of the laser resides only in the Gaussian intensity profile (hypothesis 5):

$$I_0(x, y, z) = \frac{2P}{\pi[\omega(z - l/2)]^2} e^{-\frac{2(x^2 + y^2)}{[\omega(z - l/2)]^2}} \qquad (S12)$$

where $P$ is the laser power and the function $\omega(z - l/2)$ describes the gaussian waist radius along $z$. Note that the function has been displaced by half the cuvette width $l$ along the $z$-axis to describe the focus at the cuvette center. It can be demonstrated that in free space at any $z$:

$$\iint_{-\infty}^{\infty} I_{0(x,y,z)} dx dy = P \qquad (S13)$$

As the power is conserved along $z$, when no optically active species are present, the incident intensity can be expressed as:

$$I_0(x, y, z) \approx I_0(z) = \frac{P}{\pi[\omega(z - l/2)]^2} = \frac{P}{A(z - l/2)} \tag{S14}$$

with $A(z - l/2)$ being the cross-sectional area of the propagating beam. When the sample is present, attenuation along the propagation direction $z$ due to absorption and scattering occurs according to the Beer-Lambert law (hypothesis 4):

$$I(n, z) = I_0(z)e^{-n\sigma_{ext}z} = I_0(z)\, e^{-n(\chi_{AuNP}\,\sigma_{abs} + \sigma_{sca})z} \tag{S15}$$

By substituting Equation S14 into Equation S15:

$$I(n, z) = \frac{P}{A\left(z - \frac{l}{2}\right)} e^{-n\sigma_{ext}z} \tag{S16}$$

The laser intensity can be substituted into Equation S11, expressing the infinitesimal volume as a function of $z$, $dV = A(z - l/2)\, dz$. This substitution gives

$$d\dot{Q}(n, z) = n\,\chi_{AuNP}\,\sigma_{abs}\, P e^{-n\sigma_{ext}z}\, dz \tag{S17}$$

Integration of $d\dot{Q}(n, z)$ with respect to $z$ from 0 to $l$ gives

$$\dot{Q}(n) = P\,\frac{\chi_{AuNP}\,\sigma_{abs}}{\sigma_{ext}}\,(1 - e^{-n\sigma_{ext}l}) \tag{S18}$$

The generated heat is exchanged with the environment through convection boundary conditions along the cuvette wall surfaces, then followed by conduction across the quartz walls. At the steady state the following energy balance is reached:

$$\dot{Q}(n) = \dot{Q}_{conv} + \dot{Q}_{cond} \tag{S19}$$

The terms on the right-hand side can be expanded as follows

$$\dot{Q}(n) = \Delta T \sum_i h_i A_i + \frac{kS}{L} \Delta T = \beta \Delta T \tag{S20}$$

where $h_i$ and $A_i$ are the convective heat transfer coefficients and areas of the cooling surfaces. In particular $h_i$ values were estimated from the Nusselt number correlations for vertical and horizontal planes.[59,60] $k$, $S$ and $L$ are the conductivity, transversal section, and length of the quartz cuvette wall in contact with the sample, respectively. These parameters can all be summed up in the generalized heat transfer coefficient $\beta$. The increment in temperature obtained at thermal equilibrium is therefore

$$\Delta T(n) = \frac{P}{\beta} \frac{\chi_{AuNP}\ \sigma_{abs}}{\sigma_{ext}} (1 - e^{-n\sigma_{ext}l}) \tag{S21}$$

By two simple variable changes we can express $\Delta T$ as a function of the mean inter-polymersome distance[43] $\delta = \left(\frac{4\pi n}{3}\right)^{-1/3} \Gamma\left(\frac{4}{3}\right) = \kappa n^{-1/3}$ and of the AuNP number density $n_{AuNP} = n\, \chi_{AuNP}$:

$$\Delta T(\delta) = \frac{P}{\beta}\frac{\chi_{AuNP}\ \sigma_{abs}}{\sigma_{ext}}\left[1 - e^{-\sigma_{ext}l\left(\frac{\kappa}{\delta}\right)^3}\right] \tag{S22}$$

$$\Delta T(n_{AuNP}) = \frac{P}{\beta} \frac{\chi_{AuNP}\ \sigma_{abs}}{\sigma_{ext}} \left(1 - e^{-\frac{\sigma_{ext}l}{\chi_{AuNP}} n_{AuNP}}\right) \tag{S23}$$

**Video S1.** Visual demonstration by infrared imaging of the thermoplasmonic response of a hybrid polymersome (Au@PO) suspension to 532 nm laser exposure.

**Video S2.** Live/dead confocal microscopy imaging of T98G cells under in vitro PTT treatment ($I$ = 0.8 mW µm$^{-2}$) with hybrid polymersomes. The timestamp shows the cumulative exposure time to the LSPR excitation laser (514 nm). Live cells are labeled in green, dead cells in magenta.

**Video S3.** Live/dead confocal microscopy imaging of T98G cells treated with hybrid polymersomes without LSPR laser excitation (laser OFF). The timestamp shows, for comparison, the cumulative exposure to the LSPR excitation laser (514 nm) in the laser ON case. Live cells are labeled in green, dead cells in magenta.

**Video S4.** Live/dead confocal microscopy imaging of T98G cells under in vitro PTT treatment ($I$ = 0.8 mW µm$^{-2}$) with pristine polymersomes. The timestamp shows the cumulative exposure time to the LSPR excitation laser (514 nm). Live cells are labeled in green, dead cells in magenta.